\documentclass[prd,nofootinbib,superscriptaddress]{revtex4}

\usepackage{amsmath}
\usepackage{amsfonts}

\begin{document}

\begin{flushright}
Symmetry 2 (2010) 230-271\\
$~$\\
{\small http://www.mdpi.com/2073-8994/2/1/230/ \\
doi:10.3390/sym2010230}\\
$~$\\
$~$\\
\end{flushright}
%(This article belongs to the Special Issue Feature Papers: Symmetry Concepts and Applications)

\title{Doubly-Special Relativity: Facts, Myths and Some Key Open
Issues}

\author{Giovanni AMELINO-CAMELIA}
\affiliation{Dipartimento di Fisica, Universit\`a di Roma ``La Sapienza"\\
and Sez.~Roma1 INFN, P.le A. Moro 2, 00185 Roma, Italy}

\begin{abstract}
I report,
emphasizing some key open issues and some aspects that are
particularly relevant for phenomenology,
 on the status of the development of ``doubly-special"
relativistic (``DSR") theories with both an observer-independent
high-velocity scale and an observer-independent
small-length/large-momentum scale, possibly relevant for the
Planck-scale/quantum-gravity realm. I also give a true/false
characterization of the structure of these theories. In particular,
I discuss a DSR scenario without modification of the energy-momentum
dispersion relation and without the $\kappa$-Poincar\'e Hopf
algebra, a scenario with deformed Poincar\'e symmetries which is not
a DSR scenario, some scenarios with both an invariant length scale
and an invariant velocity scale which are not DSR scenarios, and a
DSR scenario in which it is easy to verify that some observable
relativistic (but non-special-relativistic) features are insensitive
to possible nonlinear redefinitions of symmetry generators.
\end{abstract}

\maketitle

\newpage

\tableofcontents

\newpage

\section{Introduction}
The idea  of ``Doubly-Special-Relativity"
(or ``DSR") is now about 10 years old~\cite{dsr1,dsr2},
and already a few hundred papers either fully focused on it or
considered it alongside other possibilities (see, {\it e.g.},
Refs.~\cite{dsr1,dsr2,jurekdsr,gacRoxJurek,cosmodsr,leedsr,frandar,lukiedsr,jurekDSRnew,judesvisser,dsrDIRAC,dharamdsrdirac,dsrnature,dsrlodz,leeDSRprd,rossano,dsrgzk,feoli,dsrIJMPrev,dsrchak,dsrBlautKowa,rainbowDSR,maguespace,balROXher,kodadsr,jurekkodadsr,sethdsr,dsrphen,stringydsr,mg10qg5,mignemidsr,dsrlngsold,dsroriti,corgamboa,tsr,qg2DDSRfreidel,qg2DDSRoriti,lqgDSR1,lqgDSR2,gho1,gho2,gho3,gho4,leivadsr,lngsdsr,freepartyDSR,ling,konopdsr,irandsr,dsrDas,dsrSaoPaulo,dsrAurelio0607,dsrFinsl,hosse1,hosse2,dsr5Da,dsr5Db,dsrMadrid,dsrZagreb,dsrBeijing,mignemiFinsler,dsrchaiho,dsrHinter,dsrGianluca}
and references therein).
This allowed to achieve rapidly significant progress
in exploring the DSR concept, but, as inevitable for such a large
research effort over such a short time, it also affected negatively
the process of establishing a common language and common conventions.
It is clearly necessary at this point to devote some efforts to
 comparing different perspectives.
The pace of development was such that it could not be organized ``in
series", with each new result obtained by one group being metabolized
and used by other groups to produce the next significant result,
but rather ``in parallel", with research groups or clusters of
research groups obtaining
sequences of results from within one approach, and then attempting to compare
to similar sequences of results obtained by other research groups
when already a barrier of ``local dialects", intuitions
and prejudices has settled in.

It might be useful at this stage of the development of DSR
to pause and try to find some unifying features among the encouraging
results found adopting different approaches,
and confront from a perspective that might combine the
different approaches some of the most stubborn ``unsolved
issues" which often appear, although possibly differently disguised,
in all approaches.

In the first part of these notes I follow my original discussion
of Refs.~\cite{dsr1,dsr2} proposing the physics idea of a DSR theory.
And to make the concept clearer I here also characterize it in the form
of a ``true/false exercise",
whose entries are chosen on the basis of the experience
of these past few years, in which some aspects of the DSR concept
have been occasionally misunderstood.
In particular,
through the illustrative example of theories with ``canonical" spacetime noncommutativity
I discuss the possibility of a DSR
scenario without modifications of the energy-momentum dispersion
relation and without the $\kappa$-Poincar\'e Hopf algebra.
Through the illustrative example of some results originally obtained
by Fock, I discuss the possibility of a relativistic theory
with both an invariant length scale and an invariant velocity scale,
which is not a DSR
scenario. And I use some recent results on the relation between
the conserved charges and the algebraic properties of the generators of
a Hopf algebra as an example of cases in which some characteristic
relativistic (but non-special-relativistic)
features of a theory are insensitive to possible nonlinear redefinitions of
the symmetry generators.

Whenever possible I rely only on the physics concept of a DSR theory,
without advocating one or another mathematical formalism, and I stress
that this is at this stage still advisable since, although many encouraging
results have been obtained, no mathematical formalism has been fully proven
to be consistent with the DSR principles. (Actually for all of the
candidate DSR
formalisms that are under consideration the results obtained so far are not
even sufficient to rigorously exclude the presence
of a preferred frame\footnote{I shall occasionally
characterize violations of Galilei's Relativity Principle as cases
in which a ``preferred frame" emerges. Of course, even when the Relativity
Principle is violated there is a priori no reason to prefer one frame
over another, but violations of the Relativity Principle render different
inertial observers distinguishable so that a criterion to prefer
one over another is devisable.}, and it would therefore be
rather dangerous to identify the DSR concept with one or another of these
formalisms.)

When I do find useful to discuss some
aspects of the DSR concept and some of the ``open issues"
in terms of a candidate mathematical formalism,
 I shall primarily resort to the mathematics
of Hopf-algebra spacetime symmetries,
which  at present appear to provide the most promising candidate for
a formalism able to accommodate the DSR principles.
But I shall also (although more briefly) comment
on other candidate DSR formalisms, and in doing so I shall
stress the need to characterize these different approaches in
terms of (at least in-principle) observable features. It is not implausible
that some of these formally different approaches actually describe
the same DSR physical theory.

I also offer some remarks (mainly in Section 6) on
DSR phenomenology, primarily with the objective of showing
that (in spite of the present, rather preliminary, stage of developments
of DSR formalisms) we can already establish rather robustly
certain general features of this phenomenology.

\section{Relativity, Doubly Special}

\subsection{Motivation}
My proposal~\cite{dsr1,dsr2} of the doubly-special-relativity scenario
intended to provide
an alternative perspective on the studies of
quantum-gravity-induced Planck-scale departures
from Lorentz symmetry which had been presented in numerous
articles (see, {\it e.g.},
Refs.~\cite{grbgac,gampul,kifune,mexweave,biller,ita,aus,gactp})
between 1997 and 2000. These studies were advocating  Planck-scale
modifications of the energy-momentum dispersion relation, usually
of the form\footnote{The relevant literature focuses on ultra-high energy
(but still ``sub-Plankian") particles, so that $E\simeq p$ (and in any case one
never expects experimental sensitivities that would render meaningful correction terms
of the type, say, $\eta L_p^n m^2 E^n$ at ultra-high energies),
so it is customary to write
interchangeably $E^2 \simeq p^2+m^2+\eta L_p^n p^2 E^n$
and $E^2 \simeq p^2+m^2+\eta L_p^n E^{n+2}$. I am choosing to
write $E^2 \simeq p^2+m^2+\eta L_p^n p^2 E^n$ so that I can safely treat $m$
as the rest energy of the particle.} $E^2=p^2+m^2+\eta L_p^n p^2 E^n +O(L_p^{n+1}E^{n+3})$,
on the basis of preliminary findings in the analysis of several
formalisms in use in Planck-scale/quantum-gravity theoretical physics. The complexity of the
formalisms is such that very little else was known about their
physical consequences, but the evidence of a modification of the
dispersion relation was becoming robust. In all of the relevant
papers it had been assumed that such modifications of the dispersion
relation would amount to a breakdown of Lorentz symmetry, with
associated emergence of a preferred class of inertial observers
(usually identified with the natural observer of the cosmic
microwave background radiation). I was intrigued by a striking
analogy between these developments and the developments which led
to the emergence of Special Relativity, now more than a century
ago. In Galilei Relativity there is no observer-independent scale,
and in fact the energy-momentum relation is written as
$E=p^2/(2m)$. When it became clear that electromagnetic phenomena could be
formalized in terms of  Maxwell
equations, the fact that those equations involve a
fundamental velocity scale appeared to require the introduction of
a preferred class of inertial observers. But in the end we figured
out that the situation was not demanding the introduction of a
preferred frame, but rather a modification of the laws of
transformation between inertial observers. Einstein's Special
Relativity introduced the first observer-independent relativistic
scale (the velocity scale $c$), its dispersion relation takes the
form\footnote{For the majority of formulas in these notes
I am adopting conventions with $c=1$. I write $c$ explicitly only when
it seems appropriate to stress its role in a certain equation.}
 $E^2 = c^2 p^2 + c^4 m^2$ (in which $c$ plays a crucial role
for what concerns dimensional analysis), and the presence of $c$
in Maxwell's equations is now understood as a manifestation of the
necessity to deform the Galilei transformations.

I argued in Refs.~\cite{dsr1,dsr2} that it is not implausible that we
might be presently confronted with an analogous scenario. Research
in quantum gravity is increasingly providing reasons of interest
in Planck-scale modifications of the dispersion relation, of the
type mentioned above, and, while it was customary to assume that
this would amount to the introduction of a preferred class of
inertial frames (a ``quantum-gravity ether"), the proper
description of these new structures might require yet again a
modification of the laws of transformation between inertial
observers. The new transformation laws would have to be
characterized by two scales ($c$ and $L_p$) rather than the single
one ($c$) of ordinary Special Relativity.

\subsection{Defining the Concept}
The ``historical motivation" described above leads to a scenario
for Planck-scale physics which is not intrinsically equipped with
a mathematical formalism for its implementation, but still is
rather well defined. With Doubly-Special Relativity one looks for
a transition in the Relativity postulates, which should be largely
analogous to the Galilei $\rightarrow$ Einstein transition. Just
like it turned out to be necessary, in order to describe
high-velocity particles, to set aside Galilei Relativity (with its
lack of any characteristic invariant scale) and replace it with
Special Relativity (characterized by the invariant velocity scale
$c$), it is at least plausible that, in order to describe
ultra-high-energy particles, we might have to set aside Special
Relativity and replace it with a new relativity theory, a DSR,
with two characteristic invariant scales, a new
small-length/large-momentum scale in addition to the familiar
velocity scale.

A theory will be compatible with the DSR principles if there is
complete equivalence of inertial observers (Relativity Principle)
and the laws of transformation between inertial observers are
characterized by two scales, a high-velocity scale and a
high-energy/short-length scale. Since in DSR one is proposing to
modify the high-energy sector, it is safe to assume that the
present operative characterization of the velocity scale $c$ would
be preserved: $c$ is and should remain the speed of massless
low-energy particles. Only
experimental data could guide us toward the operative description
of the second invariant scale $L_{dsr}$.

These characteristics can be summarized~\cite{dsr1,dsr2} of course in the form
of some ``DSR principles".
First the statement that Galilei's relativity principle is valid:
\begin{itemize}
\item{(RP):} The laws of physics are the same in all inertial
frames (for all inertial observers); in particular, the parameters that appear in
the laws of physics take the same value in all inertial frames and, equivalently,
if two inertial observers in relative motion setup the same experimental procedure
they get exactly the same (possibly dimensionsful) numerical values for the measurement results
%%%JOC%%%%
% O misura vita media muoni a riposo (a riposo rispetto ad O) e trova risultato
% uguale ad O' che misura vita media muoni a riposo (a riposo rispetto ad O'
%
% muon lifetime puo' dipendere da, say, "posizioni stelle" se le leggi dicono che
% lifetimes depend on gravitational fields ma questo e' solo perche' nion si riesce
% a setup exactly same measurement procedure: O fa la misura di muon lifetime con le stelle
% piazzate in modo diverso rispetto a lui se lo si compara al caso di come quelle
% stelle sono piazzate rispetto ad O'
%
% DOMANDA: in che modo l'etere e' diverso dalle "stelle semifisse"?
% perche' non e' possibile in linea di principio per O' spostare l'etere e mettersi
% nelle stesse condizioni di O ?
% In linea di principio O' potrebbe spostare le stelle del suo laboratorio
% in modo da avere la stessa "configurazione relativa" di stelle in cui lavora O
%
% Nota bene: questo caso di muon proper lifetime misurata da diversi osservatori
% forse e' l'affermazione della parte attiva del PR. Ed in quel contesto emerge
% l'ostruzione verso l'introduzione dell'etere.
% Per la parte passiva del PR (se esiste) si puo' pensare di usare l'esempio
% del "decay energy threshold": la legge, say, "particle puo' decadere se E > Ma + Mb"
% non fa intervenire l'etere ma non e' un confronto tra invarianti relativistici
% ed una data particella ha energie E1 per un osservatore ed E2 per un altro osservatore
% se  E1 > Ma + Mb ed  E2 < Ma + Mb allora la particella potrebbe decadere per O1
% ma non potrebbe decadere per O2
%
\end{itemize}
Then there must be a principle giving the operative definition of the
length scale $L_{dsr}$ (or a corresponding momentum/energy/frequency scale $1/L_{dsr}$).
Since at present we have no data one can only describe the general
form of this law~\cite{dsr1,dsr2}
\begin{itemize}
\item{(La):} The laws of physics, and in particular the laws of transformation
between inertial observers,
involve a fundamental/observer-independent small (possibly Planckian)
length scale $L_{dsr}$, which
can be measured by each inertial observer following the measurement
procedure ${\cal M}_{L_{dsr}}$
\end{itemize}
And finally one must have the speed-of-light-scale\footnote{Notice that (Lb) is not really conceptually
different from the standard speed-of-light postulate of Special Relativity.
Actually the standard formulation of
speed-of-light postulate of Special Relativity is phrased redundantly.
Einstein could have adopted (Lb) and {\underline{derive}} the wavelength
independence from the absence of a fundamental length scale in Special Relativity.
By straightforward dimensional analysis one indeed concludes that
an observer-independent law
of wavelength dependence of the speed of light
is not possible in theories that do not have
an observer-independent length scale.
The fact that I adopt (Lb) {\underline{should not be interpreted}}
as suggesting that the speed of light must necessarily have
some wavelength dependence in a  relativistic theory in which there is
also an observer-independent length scale.} principle~\cite{dsr1,dsr2}:
\begin{itemize}
\item{(Lb):} The laws of physics, and in particular the laws of transformation
between inertial observers,
involve a fundamental/observer-independent velocity scale $c$, which
can be measured by each inertial observer as the speed of light
with wavelength $\lambda$
much larger than $L_{dsr}$ (more rigorously, $c$ is obtained as
the infrared $\lambda/L_{dsr} \rightarrow \infty$ limit of the speed of light)
\end{itemize}

\noindent
The postulate (La) is clearly incomplete: for the description
of the  ${\cal M}_{L_{dsr}}$  procedure for the measurement of $L_{dsr}$
 we do not have enough experimental
information to even make an educated guess. There are many physical arguments
and theoretical models that predict one or another physical role for the
Planck length, but none of these scenarios has any experimental support. It
is still plausible that the Planck length has no role
in space-time structure and kinematics
(which would render DSR research of mere academic interest).
Even assuming DSR actually does play a role in the description of Nature
it seems likely that the correct formulation of ${\cal M}_{L_{dsr}}$
would end up being different from any proposal we can contemplate
presently, while we are still lacking the needed guidance of experimental
information, but through the study of some specific examples we can already
acquire some familiarity with the new elements required by a conceptual
framework in which the Relativity Principle coexists with
observer-independent high-velocity and small-length
scales.
In light of this situation in Ref.~\cite{dsr1,dsr2}
 I considered a specific illustrative example
of the postulate (La), which was also inspired by the mentioned 1997-2000
studies of Planck-scale modification of the dispersion relation:
\begin{itemize}
\item{(La$^*$):}
The laws of physics, and in particular the laws of transformation
between inertial observers,
involve a fundamental/observer-independent small (possibly Planckian)
length scale $L_{dsr}$, which
can be measured by each inertial observer
by determining the dispersion relation for photons. This relation has
the form $E^2 - c^2 p^2 + f(E,p;L_{dsr})=0$, where
the function $f$ is the same for all inertial observers
and in particular all inertial observers agree on the
leading $L_{dsr}$ dependence
of $f$: $f(E,p;L_{dsr}) \simeq L_{dsr} c p^2 E$
\end{itemize}
In these past decade of course other examples
of measurement procedure ${\cal M}_{L_{dsr}}$
have been considered by myself and others.

\subsection{A Falsifiable Proposal}
A key objective for these notes is to expose the physics content of the DSR proposal
in ways that are unaffected by the fact that this proposal does not
at present specifiy a mathematical formalism.
Consistently with this objecties let me close
this section on the ``definition of doubly-special relativity",
by making the elementary but significant observation that, even
without specifying a mathematical formalism, the DSR idea is falsifiable.
Many alternative mathematical formalisms could be tried for DSR, and probably should be tried,
but any experimental evidence of a ``preferred frame"
would exclude DSR completely, without appeal, independently of the mathematical
formalization.
I am just stating the obvious: the idea that the laws of physics are relativistic
is of course falsifiable.
But this obvious fact carries some significance when DSR is viewed in the
broader context of the quantum-gravity literature, affected as it is by the
problem of proposals that perhaps may predict many things, but in fact, at least
as presently understood, predict nothing and/or possibly could not be falsified.

The possibility of falsifying the DSR idea  by finding evidence of a preferred frame
is actually not of mere academic (``epistemological") interest:
proposals put forward in the quantum-gravity literature that are related
to (though alternative to) DSR are motivating experimental searches of ``preferred-frame
effects" in ways that could indeed falsify the whole DSR idea if successful.
In fact, it has been recently realized (see, {\it e.g.},
Refs.~\cite{tedOLDgood,gacpion,seth}) that, when Lorentz
symmetry is broken (``preferred-frame picture") at the Planck scale, there can be significant
implications for certain decay processes. At the qualitative level
the most significant novelty would be the possibility for massless
particles to decay. Let us consider for example a photon decay
into an electron-positron pair: $\gamma \rightarrow e^+ e^-$. And
let us analyze this process using the dispersion relation
\begin{equation}
 m^2 \simeq E^2 - \vec{p}^2
+  \eta \vec{p}^2 \left({  E \over E_{p}}\right)
%\vec{p}^2
~\label{displeadbis}
\end{equation}
where $E_{p} = 1/L_{p}$. Assuming then an unmodified law of
energy-momentum conservation, one easily finds a relation between
the energy $E_\gamma$ of the incoming photon, the opening angle
$\theta$ between the outgoing electron-positron pair, and the
energy $E_+$ of the outgoing positron, which, for the region of
phase space with $m_e \ll E_\gamma \ll E_p$, takes the form
$\cos(\theta) = (A+B)/A$, with $A=E_+ (E_\gamma -E_+)$
and $B=  m_e^2 -
\eta  E_\gamma E_+ (E_\gamma -E_+)/E_p$
 ($m_e$ denotes of course the electron mass).  For $\eta < 0$
the process is still always forbidden, but for positive $\eta$
and $E_\gamma \gg (m_e^2 E_p/|\eta |)^{1/3}$ one finds that
$\cos(\theta) < 1$ in certain corresponding region of phase space.

The energy scale $(m_e^2 E_p)^{1/3} \sim 10^{13} eV $ is not too
high for astrophysics. The fact that certain observations in
astrophysics allow us to establish that photons of energies up to
$\sim 10^{14}eV$ are not unstable (at least not
noticeably unstable) could be used~\cite{tedOLDgood,seth} to set
valuable limits on $\eta$.

If following this strategy one did find photon decay then the idea
of spacetime symmetries {\underline{broken}} by Planck-scale effects would be
strongly encouraged. The opposite is true of DSR, which essentially
codifies a certain type of {\underline{deformations}} of Special Relativity.
Any theory compatible with the DSR
principle must have stable massless particles.
A threshold-energy requirement for massless-particle decay
(such as the $E_\gamma \gg (m_e^2 E_p/|\eta |)^{1/3}$ mentioned above)
cannot of course be introduced as an observer-independent law, and is therefore
incompatible with the DSR principles.
By establishing the existence of a threshold
for photon decay one could therefore indeed falsify the
DSR idea.
And more generally any theory compatible with the DSR
principle must not predict energy thresholds for the decay of particles.
In fact, one could not state observer-independently a law
setting a threshold energy
for a certain particle decay, because different observers attribute different
energy to a particle (so then the particle should be decaying according to some observers
while being stable according to other observers).

\section{More on the Concept (Some True/False Characterizations)}
The definition of DSR given in Ref.~\cite{dsr1,dsr2}, which (in order
to render this review self-contained) I here repeated
in the preceding  section, is crisp enough not to require any further
characterization. However,
partly because it does not yet come with a specification of mathematical
formalism, and partly because of some inconsistencies of terminology
that characterized the rather large DSR literature produced
in just a few years, has
occasionally generated some misinterpretations.
This motivates the redundant task to which I devote the present section:
drawing from the experience of this past decade, I explicitly comment on
some possible misconceptions concerning DSR.

\subsection{Most Studied DSR Scenarios and Inequivalence to Special Relativity}
The  DSR concept, as introduced
in Refs.~\cite{dsr1,dsr2}, is very clearly
 {\underline{alternative}} to Special Relativity.
How could it then be possible that some authors
(see, {\it e.g.}, Ref.~\cite{irandsr}) have instead
naively argued the equivalence?

What could happen
is that some authors may propose a certain formalization of a DSR theory
and then other authors show that the attempt was unsuccessful,
meaning that the proposed formalization, while intended as a DSR candidate,
actually turns out to be simply special relativistic.
However,
the cases that apparently are of this sort in the DSR literature
(see, {\it e.g.}, in Ref.~\cite{irandsr})
have actually nothing to do with establishing whether or not
one type of another of relativistic properties is found in a
certain theory proposal, and rather are due to
the mistake of reasoning on the basis of formal analogies rather
than in terms of physical/observable/operatively-well-defined concepts.

The core misconceptions that generated claims of equivalence
of some candidate DSR proposal with ordinary special relativity
originates from the role played in these candidate DSR proposals
by (one type or another, see later) modifications of the commutators
of Poincar\'e generators.
There is a certain (obviously incorrect, but surprisingly popular)
frame of mind among some theorists
according to which ``finding a map between two theories establishes their equivalence".
It is difficult for me to even pretend to
put myself in that frame of mind, considering that,
even just remaining within relativistic theories, one easily finds ``some maps"
that can convert Galileian relativity into Minkowski-space special relativity,
and in turn  Minkowski-space special relativity into deSitter-space
special relativity. And specifically a map of type very similar to the one considered
for example in Ref.~\cite{irandsr} from a pseudo-DSR perspective
can be used to convert
the Galilei Lie algebra
for 2D spacetimes
into the Poincar\'e Lie algebra for 2D spacetimes.

However,
rather than articulating in detail a critique of this ``the map is the answer"
methodology, I suppose it is more useful if I briefly characterize here two
(of the many possible)
scenarios for the formalization of the DSR concept and comment
on the futility of seeking maps that would convert some symbols
on the DSR side into special-relativistic symbols.

\subsubsection{The DSRa scenario: nonlinear representations of the Poincar\'e Lie group}
The most studied ``toy-model scenarios" (see related comments in the section on
phenomenology) for DSR essentially assume that the DSR principles
could be implemented by seeking corresponding nonlinear representations
of the standard Poincar\'e Lie group.
The action of the generators of the
 Poincar\'e Lie group
 on such a representation is such that (when specified on that representation)
 one could effectively describe the Poincar\'e symmetries in terms of
 some deformed commutators between the generators.

So this is a case where the new observer-independent scale of DSR
characterizes representations of a still classical/undeformed
 Poincar\'e Lie group.
But one can effectively handle these proposals by
replacing the standard Poincar\'e generators
with new generators, specialized to the nonlinear
representation: $ T_a \rightarrow T_{dsr}^a$.
Of course (since indeed one is merely adotping a possible
scheme of analysis of a nonlinear representations of
the standard Poincar\'e Lie group)
it is easy to find in these instances some nonlinear
map ${\cal F}$ such that $ T_a = {\cal F}(\{T_{dsr}^b\})$.

Some authors
(see, {\it e.g.}, in Ref.~\cite{irandsr})
view the availability of such
a map ${\cal F}$, converting the ``effective generators" $T_{dsr}^a$
of the DSR picture
(specialized to a nonlinear representation) into ordinary
generators $ T_a = {\cal F}(\{T_{dsr}^b\})$,
as evidence that the relevant DSR scenarios should actually
be viewed as a description of completely standard physics within
ordinary special relativity.
But of course this is extremely naive since the case with nonlinear representations
characterized by an observer-independent short-distance scale and the case
without such an observer-independent scale produce different physics,
different observable phenomena, and can be experimentally distinguished from the
ordinarily special-relativistic case.

\subsubsection{The DSRb scenario: Poincar\'e-like Hopf-algebra spacetime symmetrie}
Perhaps the most intriguing among the scenarios that are being
considered as candidates for a DSR theory
are scenarios based on the mathematics of Poincar\'e-like Hopf algebras.
I shall describe other aspects of this scenario later in these notes.
In this subsection I just want to observe that also for the Hopf-algebra
case the search of ``some nonlinear maps from the DSR candidate to ordinary special
relativity"
appears to be futile, but for reasons that amusingly
are complementary to the ones discussed in the previous subseciton for another
popular candidate DSR scenario.

In this Hopf-algebra case
dogmatic followers of ``map is the answer" methodology
have missed
a key aspect of Hopf algebras: a satisfactory characterization
of these algebras must describe for the generators not only
the commutators
but also the so-called cocommutators (the so-called
coproduct rules), which affect the law of action of generators on products
of functions. This is related to the fact that
the classic application of Hopf-algebra symmetries is to frameowrks such
that (unlike the ones treatable with Lie algebras)
the action of symmetry transformations on products
of functions is not obtainable by standard application of Leibniz rule
to the action on a single function (as I shall discuss in more detail later in these notes).
This fits naturally with the properties of
some structures used in Planck-scale research, such as noncommutative spacetimes
(since the action of operators on products of functions of the noncommutative
spacetime coordinates is naturally not subject to Leibniz rule).
Nonlinear redefinitions of generators for Hopf algebras are admitted,
just because they are futile: the form of the commutators of course changes
but there is a corresponding change in the cocommutators, and the combined effect
amounts to no change at all for the overall description of the symmetries.
In Refs.~\cite{kappanoether,thetanoether}
this result, which can be shown
already at the abstract algebra level, was worked out explicitly, for
two nonlinear redefinitions of two examples of Hopf algebra deformations
of the Poincar\'e algebra, finding that one obtains the same
conserved charges independently of whether one analyzes
the Hopf-algebra symmetries in terms of one set of generators or a nonlinear
redefinition of that given set of generators. Essentially this comes from the
fact that the Noether analysis of a theory with a given Lagrangian density
will combine aspects described in terms of the action of generators on
single functions and aspects which instead concern the action
on product of functions.

\subsubsection{Implications of the nonlinear maps for the action of generators}
Let me close this subsection on the use of nonlinear maps in some candidate DSR
frameworks with an observation that may apply to a rather broad spectrum
of scenarios: when a map from one set of generators to another is analyzed
one should not overlook the implications for the rule of action of
those generators. For my purposes it suffices to raise this issue
within a rather abstract/formal scheme of analysis, working in leading order
in a deformation length scale $\lambda$, and focusing on the
simplest case of the relativistic description of massless classical particles
in 1+1D spacetimes.
Let me start by introducing standard translation
generators $\Pi,\Omega$
\begin{equation}
[x,\Pi]=1~,~~[t,\Omega]=1~,~~[x,\Omega]=0~,~~[t,\Pi]=0~,~~[\Omega,\Pi]=0~. \label{heiseJJJ}
\end{equation}
A deformation of relativistic symmetries of DSR type may be obtained
by then adopting a non-standard representation of the boost generator
\begin{equation}
N=x(\Omega-2 \lambda \Omega \Pi) + t (\Pi+\lambda \Omega^2)
~, \label{boostomega}
\end{equation}
from which it follows that
\begin{equation}
[N,\Omega]=\Pi+\lambda\Omega^2~,~~[N,\Pi]=\Omega-2 \lambda \Omega \Pi
~, \label{deformedalgebra}
\end{equation}
which is a deformation of the commutators of the 1+1D Poincar\'e algebra,
with mass Casimir ${\cal C}_{\Omega,\Pi} = \Omega^2 - \Pi^2 - 2 \lambda \Omega^2 \Pi$.
Among the implications of this deformation most striking is the fact that
the velocity of massless classical particles is not 1. This can be formally
established by noticing that for an on-shell particle
the generators take numerical value (conserved charges),
and then differentiating (\ref{boostomega}) in $dt$ one obtains
\begin{equation}
\frac{dx}{dt} \simeq 1+ 2 \lambda \Omega_0
~, \nonumber
\end{equation}
where I denoted by $\Omega_0$ the $\Omega$-charge (energy) of the particle
and also used the massless-shell condition inferred from the casimir ${\cal C}_{\Omega,\Pi}$.

Some authors are (erroneously) intrigued by the fact that the nontriviality (\ref{deformedalgebra})
of the symmetry-algebra relations can be removed by adopting the following
nonlinear map
\begin{equation}
\epsilon = \Omega ~,~~\pi = \Pi+\lambda \Omega^2
~, \nonumber
\end{equation}
which indeed accoomplishes the task of producing from (\ref{deformedalgebra}) the following outcome
\begin{equation}
[N,\epsilon]=\pi~,~~[N,\pi]=\epsilon
~. \nonumber
\end{equation}
But it is crucial for such considerations not to miss that
upon adopting $\epsilon,\pi$ such that $\epsilon = \Omega ~,~~\pi = \Pi+\lambda \Omega^2$
it then follows from (\ref{heiseJJJ}) that
\begin{equation}
[x,\pi]=
1~,~~[t,\epsilon]=1~,~~[x,\epsilon]=0~,~~[t,\pi]= 2 \lambda \epsilon
~. \nonumber
\end{equation}
Therefore one has accomplished a trivialization of the algebraic (commutator) relation
among Poincare-sector generators
(with mass Casimir ${\cal C}_{\epsilon,\pi} = \epsilon^2 - \pi^2$)
only at the cost of introducing a nontrivial
rule of action of the ``would-be-translation generators" $\epsilon,\pi$.
It is then obvious that this nontriviality of the action of $\epsilon,\pi$ imposes
a nontrivial representation of boosts, in spite of the fact that
the Poincare-sector commutators are undeformed:
\begin{equation}
N=x (\epsilon - 2 \lambda \epsilon \pi) + t \pi
~. \nonumber
\end{equation}
And if one derives again formally the velocity of massless-shell classical particles
the result is
\begin{equation}
\frac{dx}{dt} \simeq 1 + 2 \lambda \epsilon \simeq 1+ 2 \lambda \Omega
 ~,
 \nonumber
\end{equation}
{\it i.e.} the velocity law is insensitive to the nonlinear map.
This is an example of a nonlinear map which connects two different theories
({\it e.g.} the deBroglie relation implicit in these formal manipulations is
different on the two sides of the map) which however have in common some
nontrivial predictions, such as the one for the speed of massless particles.

\subsection{Not Necessarily Involving the $\kappa$-Poincar\'e Hopf Algebra}
The formalism of Hopf algebras which I just used to illustrate the weakness
of some ``equivalence claims" found in the literature, is also
at present the most promising opportunity
to find a formalism compatible with the DSR principles.
However, it would be dangerous to identify the DSR idea
with the mathematics of Hopf algebras,
since at present we still know very little about the implications
of Hopf-algebra spacetime symmetries for observable properties
of a theory, and of course the DSR concept refers to the observables
of a theory and their properties under change of inertial observer.

Most attempts to construct a DSR theory using Hopf algebras focus on
the so-called $\kappa$-Poincar\'e Hopf algebra~\cite{lukie91e92,majrue,kpoinap}, but if indeed
Hopf algebras prove to be usable in DSR theories then other Hopf algebras
might well also be considered. I will mention in various points of these
notes the case of the twisted Hopf algebras of symmetries of
observer-independent canonical  noncommutative spacetime,
which appears to be equally promising.
And of course it is at present fully legitimate
to look for realizations of the DSR concept
that do not involve Hopf algebras.

Even if Hopf algebras do eventually turn out to be usable in
the construction of DSR theories it would be inappropriate
to identify the mathematics of (some) Hopf algebras with
the DSR physics concept.
This observation becomes obvious if one considers,
for example,
the nature of the works devoted to the $\kappa$-Poincar\'e
Hopf algebra before the proposal of the DSR concept.
In the pre-DSR $\kappa$-Poincar\'e literature one finds
some warnings~\cite{kpoinnogroup} against
attempts to integrate the boost generators to obtain
a candidate for
finite boosts, which appeared to lead only to
a ``quasi-group"~\cite{kpoinnogroup,batalin} structure,
of unestablished applicability in physics.
Similarly, the law of energy-momentum conservation which is
advocated by some $\kappa$-Poincar\'e experts~\cite{lukiedsr}
is incompatible
with the DSR principles~\cite{mg10qg5}.
These challenges of course do not imply that we should necessarily
exclude the use of the $\kappa$-Poincar\'e
formalism in attempts to construct DSR theories; it only suggests
that such attempts should rely on some carefully devised
interpretation of the symbolism.

\subsection{Not Any Deformation, but A Certain Class of Deformations of Special Relativity}
While, as I just stressed, it is clearly too early to associate with
the DSR concept a specific mathematical formalism, it would also be
disastrous to gradually transform (e.g., by gradual modifications of
the definition of a DSR theory) the DSR concept into a large
umbrella covering all scenarios for a length scale to enter
spacetime symmetries. It is for this reason that I chose to do the
redundant exercise of repeating here (in Section 2) exactly the
definition of DSR theory originally given in Refs.~\cite{dsr1,dsr2}.
It actually did happen over this past decade that occasionally the
DSR proposal was confused as the proposal that we should have some
deformation of Special Relativity, with the idea that any
deformation would be DSR-acceptable. The definition given in
Refs.~\cite{dsr1,dsr2} (here repeated in Section~2) provides a
physics picture which amounts instead to a rather specific class of
deformations of Special Relativity, alternative to other
possibilities.

It is of course very easy to show
 that one could ``deform" Special Relativity in ways
that are not compatible with the DSR concept.
One should think for example of deSitter relativity,
which is of course a deformation of Special Relativity by the scale of curvature.
And it is interesting that one could actually stumble upon deSitter space
just in doing the exercise of looking for ``some deformation" of Special Relativity.
This actually happened: Fock, in an appendix of Ref.~\cite{fockbook},
 explores the role that each of the postulated structures of Special
Relativity plays in constraining the mathematics of the Special-Relativity
framework. Unsurprisingly by removing one of the in-principle structures
Fock obtains De Sitter spacetime\footnote{Fock
does not appear to realize he has actually landed on De Sitter, but does handle
all formulas correctly.} instead of Minkowski spacetime.
De Sitter spacetime is a deformation of Minkowski spacetime
and the De Sitter algebra is a deformation of the Poincar\'e algebra
(mediated by the Inonu-Wigner contraction procedure), but of course
these standard formalisms do not provide examples of DSR theories.
Indeed De Sitter spacetime is a deformation of Minkowski spacetime
by a {\underline{long}}-distance scale (Minkowski obtained from De Sitter
as the deformation length scale is sent to infinity),
whereas one of the requirements for a DSR theory
is that the deformation scale be a short-distance (high-energy) scale
(Special Relativity obtained from DSR as the
deformation length scale is sent to zero).

Another example worth mentioning is the one of proposals in which one introduces
a maximum-acceleration scale. Since already
in Special Relativity acceleration
is an invariant, such proposals do not {\it a priori}
require a DSR formulation.
Indeed often these proposals involve formalisms able to handle at once
both inertial observers and essentially Rindler observers (of course acceleration
changes when going from an inertial frame to a Rindler frame).
At least in a generalized sense these are also ``deformations" of Special
Relativity, but typically they do not require a modification of the
laws of transformation between inertial observers.

As these two examples illustrate clearly, not any ``deformation" of
Special Relativity provides a realization of the DSR concept. I
guess the confusion some authors have on this point originates from
a recent fashion to use ``Deformed Special Relativity" as an
equivalent name for Doubly-Special Relativity. Of course, there is
no content in a name and one might consider it a free choice of the
writer, but there are cases in which a certain choice of name may
induce confusion, and this is certainly one of those cases, since
``Deformed Special Relativity" invites a naive interpretation of the
type ``any deformation of Special Relativity will do". And an
additional source of confusion originates from the fact that in the
literature, already before the proposal of the Doubly-Special
Relativity idea, there was a research programme named ``deformed
special relativity" (see, {\it e.g.,}, Ref.~\cite{cardone} and
references therein), which pursues physics motivation and physics
objectives that are completely different from the ones of
Doubly-Special Relativity.

\subsection{A Physics Picture Leading to DSR and the Possibility
of DSR Approximate Symmetries} One other point which I stressed in
Refs.~\cite{dsr1,dsr2} and has been largely ignored concerns the
possibility that ``DSR symmetries" might actually be only
approximate symmetries, even within the regime of physics
observations where they do (hopefully) turn out to be relevant. In
order to clarify this point let me first discuss a certain ``vision"
for the structure of spacetime at different scales.

While at present the status of development of research in quantum
gravity allows of course some sort of ``maximum freedom" to imagine
the structure of spacetime at different distance scales, there are
arguments (especially within the ``emergent gravity" literature, but
also beyond it) that would provide support for the following
picture: (I) at superPlanckian distance scales the only proper
description of spacetime degrees of freedom should be ``strongly
quantum", so much so that no meaningful concept of spacetime
coordinates and continuous spacetime symmetries could be introduced,
(II) then at subPlankian but nearly Planckian scales some sort of
intelligible geometry of spacetime might emerge, possibly allowing
the introduction of some spacetime coordinates, but the coordinates
might well be ``quantum coordinates" ({\it e.g.,} noncommutative
coordinates) and the symmetries of this type of spacetime geometry
might well be DSR symmetries, (III) then finally in the infrared our
familiar smooth classical description of spacetime becomes a
sufficiently accurate description.

Besides providing a logical scheme for the emergence of DSR symmetries, this picture
also explains in which sense I could argue that DSR symmetries might well be
only approximate symmetries, even in the regime where they are relevant.
Most authors (including myself) working on DSR insist on getting formulas that
make sense all the way up to infinite particle energies. There is nothing wrong
with that, but it might be too restrictive a criterion. If DSR is relevant
only at energy scales that are subPlankian (but nearly Planckian), and if we actually
expect to give up an intelligible picture of spacetime and its symmetries above the
Plank scale, then perhaps we should be open to the possibility of using mathematics
that provides an acceptable (closed) logical picture of DSR only at leading
order (or some finite order) in the expansion of formulas in powers
of the Planck length.

What I mean by this will be made somewhat clearer later in this notes,
when I set up the discussion of DSR phenomenology. There a candidate
DSR test theory
will be based on
formulas proposed exclusively as leading-order formulas.

\subsection{Not Any Fundamental Length Scale}
Inadequate description/investigation of the laws of transformation
between inertial observers is, as stressed in part in the previous subsections, a
very serious limitation for an analysis being proposed
as a DSR study. However, there is worse: there are (fortunately only very few)
self-proclaimed ``DSR studies" which are only structured at the level of proposing
some ``fundamental length/energy scale", without any effort to establish whether
this scale is of the type required by the DSR concept (in particular not providing
any analysis of  the role that this ``fundamental scale"
does or does not have in the laws of transformation between observers).

Unfortunately, even outside the DSR literature,
it is not uncommon to find in the physics
literature a rather sloppy use of terms such as ``fundamental
scale". In particular, ``fundamental scales" are often discussed
as if they were all naturally described within a single category.
For DSR research it is instead rather important that these
concepts be handled carefully.

In this respect the DSR concept is conveniently characterized through the
presence of a short-distance (/high-energy) scale which is ``relativistically fundamental"
in the same sense already familiar for the role of the scale $c$ in Special Relativity.

There are of course scales that are no less fundamental, but have properties
that are very different from the ones of $c$ in Special Relativity.
Obvious examples are scales like the rest energy of the electron, which is
of course a ``fundamental" scale of Nature, but is relativistically trivial
(a rest-frame property).
Just a bit less obvious are cases like the one of the
 quantum-mechanics scale $\hbar$.
Space-rotation symmetry is a classical continuous symmetry. One
might, at first sight, be skeptical that some laws
(quantum-mechanics laws) that discretize angular momentum could
enjoy the continuous space-rotation symmetry, but more careful
reasoning~\cite{areaNEWpap} will quickly lead to the conclusion
that there is no {\it a priori} contradiction between
discretization and a continuous symmetry. In fact, the type of
discretization of angular momentum which emerges in ordinary
non-relativistic quantum mechanics is fully consistent with
classical space-rotation symmetry. All the measurements that
quantum mechanics still allows (a subset of the measurements
allowed in classical mechanics) are still subject to the rules
imposed by rotation symmetry. Certain measurements that are
allowed in classical mechanics are no longer allowed in quantum
mechanics, but of course those measurements cannot be used to
characterize rotation symmetry (they are not measurements in which
rotation symmetry fails, they are just measurements which cannot
be done).
A more detailed discussion of this point can be found in
Ref.~\cite{areaNEWpap}. Essentially one finds that $\hbar$ is not
a scale pertaining to the structure of the rotation
transformations. The rotation transformations can be described
without any reference to the scale $\hbar$. The scale $\hbar$
sets, for example, the minimum non-zero value of angular momentum
($L^2_{min}=3 \hbar^2/4$), but this is done in a way that does not
require modification of the action of rotation transformations.
Galilei boosts are instead genuinely inconsistent with the
introduction of $c$ as observer-independent speed of massless
particles (and maximum velocity attainable by massive particles).
Lorentz/Poincar\'e transformations are genuinely different from Galilei
transformations, and the scale $c$ appears in the description
of the action of Lorentz/Poincar\'e generators
(it is indeed a scale of ``deformation" of the Galilei transformations).

Both $\hbar$ and $c$ are fundamental scales that establish
properties of the results of the measurements of certain
observables. In particular, $\hbar$ sets the minimum non-zero
value of angular momentum and $c$ sets the maximum value of speed.
But $\hbar$ has no role in the structure of the transformation
rules between observers, whereas the structure of the
transformation rules between observers is affected by $c$. I am
describing $c$ as a ``relativistically fundamental"
scale,
whereas $\hbar$ is a ``relativistically trivial" scale,
a fundamental scale that does not affect the
transformation rules between observers.

A characterizing feature of the DSR proposal is that there should
be one more scale playing a role analogous to $c$, and it should be
a short-distance/high-energy scale.

One can try to introduce the Planck scale in analogy with $\hbar$
rather than with $c$. In particular, Snyder looked\footnote{In his renowned
paper in Ref.~\cite{snyder}, Snyder explicitly states its objectives
as the ones of introducing spacetime noncommutativity
in a way that is compatible
with Special Relativity. From the tone of the paper it appears
that Snyder was
certain to have succeeded, but actually it might be necessary
to reexamine  his proposal with the more powerful tools of
symmetry analysis that have been
recently developed.
It appears to be not unlikely that some of Snyder's conclusions
might be based on a incorrect description of energy-momentum~\cite{mg10qg5}.
From a physics perspective it remains possible that,
while Snyder was proposing an ordinarily special-relativistic framework (and thought
he succeeded), a more careful analysis of the mathematics of the Snyder noncommutative
geometry might lead to structures suitable for a DSR theory~\cite{mg10qg5}.}
for a theory~\cite{snyder} in which spacetime coordinates would not commute, but
Lorentz transformations would remain unmodified by the new
commutation relations attributed to the coordinates. Such a
scenario would of course {\underline{not}} be a DSR scenario.

In closing this subsection I should comment on one more type of
fundamental constants. The electron-mass scale $m_e$, the quantum-mechanics
scale $\hbar$ and the (infrared-limit-of-the-) speed-of-light scale $c$ are
different types of fundamental constants, whose operative
meaning is of course given through the measurements of certain corresponding
observables. Another type of fundamental constants are
the ``coupling constants". For example, in our present description of
physics the gravitational coupling $G$ is a fundamental constant.
It does not impose constraints on the measurements of a specific
observable, but it governs the laws of dynamics for certain
combinations of observables. Also $G$ is observer independent,
although a careful analysis (which goes beyond the scopes of this
note) is needed to fully characterize this type of
fundamental scales. One can define $G$ operatively through the
measurement of static force between planets. In modern language
this amounts to stating that we could define $G$ operatively as
the low-energy limit of the gravitational coupling constant. All
observers would find the same value for this (dimensionful!)
constant.

This last remark on the nature of the fundamental constant $G$ is
particularly important for DSR theories. In our present
description of physics the Planck length $L_p$ is just the square root
 of $G$, rescaled through $\hbar$ and $c$. The idea of
changing the status of $G$ ({\it i.e.} $L_p$) from the one of
fundamental coupling scale to the one of relativistic fundamental
scale might have deep implications~\cite{dsr1,dsr2,dsrIJMPrev}.

\subsection{And the Same Type of Scale may or may not be DSR Compatible}
In the previous subsection I stressed the difference between
scales that are potentially significant from a relativistic perspective
(such as $c$)
and scales that are not (such as $\hbar$ and the electron rest energy).
It is important to also stress that even when dealing with a scale
that is potentially significant from a relativistic perspective,
without a description/analysis of the laws of transformation between
inertial observers any claim of relevance from a DSR perspective is futile.

Consider for example the possibility of a ``minimum wavelength".
This is clearly a type of proposal that is potentially significant
from a relativistic perspective, and indeed
one of the objectives of DSR research is to find a meaningful framework
for the implementation of a ``minimum wavelength principle" in
an {\underline{observer-independent}} manner.
However, the proposal of a minimum wavelength does not in itself
provide us with a DSR proposal. In particular, the minimum-wavelength bound
could be {\underline{observer dependent}}. A calculation
within a given theory providing evidence of a minimum-wavelength bound
would of course not suffice to qualify the relevant theory as a DSR theory.
One should at least also analyze what type of laws of transformation
between observers apply in the relevant theory and verify that the minimum-wavelength
bound (both the presence of a bound and the value of the bound) is
observer-independent.

Similar considerations apply for proposals of modified
energy-momentum dispersion relations and modified
wavelength-momentum (deBroglie) relations. It is well understood
that some frameworks in which one finds modified dispersion
relations are actually not compatible with the DSR principles: in
the relevant frameworks the modification of the dispersion relation
is observer dependent (at least because of observer dependence of
the scale of modification of the dispersion relation), and
Poincar\'e symmetry is actually broken. Analogously in
quantum-gravity scenarios (see, {\it e.g.,}
Refs.~\cite{kempmang,dadebro}) in which one adopts a modified
relation between momentum and wavelength (typically such that
momentum is still allowed to go up to infinity but there is a finite
minimum value of wavelength), before concluding in favour or against
compatibility with the DSR principles an analysis of the laws of
transformation between observers is of course necessary. One could
imagine such a modification of the relation between momentum and
wavelength to be a manifestation of a breakdown of Poincar\'e
symmetry, but there is also no in-principle obstruction for trying
to implement it in a DSR-compatible way.

\section{Spacetime ``Quantization" and Relativistic Paradoxes}
Let me now comment on some
issues concerning ``spacetime quantization and relativistic paradoxes" in DSR,
from a perspective which in part also contributes to a true/false characterization
of the DSR concept: during this first decade of DSR research several authors
appear to have been devoting a significant effort toward showing
that there are formulations of DSR compatible with
a classical-spacetime picture and such that no ``relativistic paradox" is produced,
but from the perspective I shall here advocate it appears that these efforts
are based on incorrect assumptions.

At least on the basis of experience (let me postpone to another occasion the logical
grounds for this) we should actually assume that
the transition from a given relativistic theory to the next
brings along (apparent) ``paradoxes". This is simply because the relativistic
framework in which the laws of physics are formulated affects deeply
the nature of the observables that appear in those laws, first by characterizing
the objectivity of physical processes (not withstandding a possible subectivity, {\it i.e.}
observer dependence, of the actual numerical results of measurement of some observables
by different observers of the same system),
and then in turn affecting our intuitive perception of those processes.

When Galileo introduced his relativistic theory, with anecdotal fame proportional
to the fact that the some of the guardians
of the previous (non-)relativistic theory where at the Vatican,
a major ``paradox" was that within Galieleian relativity
(adopting modern contextualization) a basketball bouncing on a moving
ship would keep landing on the same point of the floor of the ship:
%how come the ship
%does not manage to travel ahead of the ball while the ball is in the air?
how could the basketball keep up with the ship when it is in the
air, without any contact with the ship, and there is therefore no
way for the ship to carry it along? The answer was perhaps very
deep, but not truly paradoxical: it could, and it does.

With the advent of Einstenian relativity we were faced with the
novelty of relative time and length, producing (among many other
paradoxes) the famous ``twin paradox" and ``pole-in-a-barn paradox".
We now know for a fact that some twins (at least twin particles) do
age differently (if their histories of accelerations are different),
but I sometime wonder whether Galileo could have come to terms with
it, had he managed to stumble upon corresponding experimental
evidence.

The fact that such a wide spectrum of tentative formalizations of DSR are being considered
(and many more could clearly be considered) does not allow me to claim that necessarily
the correct formulation(s) should have spacetime quantization and ``relativistic paradoxes".
But on the basis of  analyses of some  of the candidate DSR formalizations
and on the basis of the history of the advent of previous relativistic theories
I conjecture that this will be the case. Chances are there is actually no room for DSR
in the laws of Nature, but if there was and it came without affecting deeply our intuition
(thereby producing apparent paradoxes) it would be a major let down.

If I aimed  for full generality I could not possibly go beyond
intuition and desires, but I believe it is worth elaborating on
these issues in at least one explicit context. And from this
perspective it is particularly intriguing to contemplate DSR
scenarios in which the speed of massless particles ({\it e.g.,}
photons) acquires a dependence on energy/wavelength. Because of the
role that the speed of light has played in the history of physics it
is certainly noteworthy that the DSR framework, by introducing an
observer-independent energy/length scale, could provide room for a
relativistic (observer-independent) law introducing a dependence on
energy/wavelength for the speed of light. And indeed a large part of
the DSR literature has considered this possibility, both because  it
is intriguing conceptually and because tests of energy dependence of
the speed of photons are improving quickly, so that there could be
opportunities for test/falsification of such DSR scenarios. But of
course one can have DSR relativistic theories without such energy
dependence of the speed of photons (by introducing the additional
relativistic scale in other ways, such as, {\it e.g.,}, a ``minimum
length uncertainty" principle), and in deciding how much effort
should be directed toward DSR scenarios with this energy
dependence of the speed of photons and how much should be directed
toward other DSR scenarios is may be valuable to fully appreciate
the logical implications of an observer-independent law for the
energy dependence of the speed of photons. In this section I will
expose some of the relativistic paradoxes that are produced by an
observer-independent law for the energy dependence of the speed of
photons. At the level of contribution to the intuition of theorists
these ``paradoxes" can actually go both ways: some readers will find
in them reason to focus on other DSR scenarios, others will perceive
them as just the right ``amount of paradoxality" that a truly new
relativistic theory should introduce. But the technical implications
of the observations reported in this section are unequivocal: it is
futile to seek a classical-spacetime formulation of DSR theories
with energy dependence of the speed of photons.

For the purposes of this section it suffices for me to contemplate a DSR scenario
with the following properties:
 \begin{itemize}
\item{(i)} for the reminder of this section it will be assumed that the
speed of a ``ultrarelativistic" ($E\gg m$) particle of mass $m$ and energy $E$ is given by
\begin{equation}
v(E) \simeq 1 - \frac{m^2}{2 E^2} + \lambda E ~ \label{velPARADO}
\end{equation}
\item{(ii)} and also for the reminder of this section it will be assumed that the
laws of transformation between inertial observers are deformed in such a way to leave
invariant the relationship
\begin{equation}
E^2 \simeq p^2 +m^2 + \lambda p^2 E~\label{mdrPARADO}
\end{equation}
 and in particular
the generators of boosts
take the form
\begin{equation}
{\cal B}_j \simeq i p_j \frac{\partial}{\partial E}+ i \left(E +
\frac{\lambda}{2} {\vec p}^2 + \lambda E^2 \right)
\frac{\partial}{\partial p_j} -i \lambda p_j  \left(p_k
\frac{\partial}{\partial p_k} \right )~ \label{dsr1boostPARADO}
\end{equation}
\end{itemize}

In some DSR studies these 3 relations have been tentatively treated as
parts of a consistent picture (different authors considering
different frameworks, but sharing these 3 features)
on the basis of the mentioned logical link
between (\ref{mdrPARADO}) and (\ref{dsr1boostPARADO})
({\it i.e.} the observation that adopting (\ref{dsr1boostPARADO}) one
finds that (\ref{mdrPARADO}) is a relativistically invariant relation)
and observing that (\ref{velPARADO})
follows from (\ref{mdrPARADO}) if one preserves the validity
of the (group-)velocity relation $v=dE/dp$.

However, as I shall show in the reminder of this section,
(\ref{velPARADO}) and (\ref{dsr1boostPARADO}) combine to
lead to a description of distances and time intervals which
is ``paradoxical", at least in the sense that it requires a very specific
(and apparently rather ``virulent") non-classical/quantum structure for
the description of spacetime.
In light of these observations a legitimate choice is the one
disregarding this possibility, and therefore focus DSR research on some
of the many alternative possibilities.
But it is also interesting to insist on the validity of these 3 ingredients
and verify whether one actually runs into direct conflict with
any established experimental facts. This is an exercise to which I devoted
some effort, finding that there appears to be no such conflict with experiment,
but rather the (scary but intriguing) need of revising very deeply
our conceptualization of spacetime.

Through this exercise one also ends up putting in focus a plausible
scenario for DSR, according to which one performs a genuine deformation
of the special-relativistic properties of energy-momentum space, but
one modifies the special-relativistic description of spacetime
in a more virulent fashion, which might be not properly
described by the term ``deformation" (usually intended as a relatively soft
modification, smoothly recovering the original theory as the deformation
parameter is removed).

\subsection{Spacetime Fuzziness for Classical Particles}
The fact that assuming (\ref{velPARADO}) and (\ref{dsr1boostPARADO})
one must immedietaly renounce to a classical spacetime picture
was first established in Ref.~\cite{dsrIJMPrev}.
The argument relies~\cite{dsrIJMPrev} on contemplating
two classical massive particles, with different
masses $m_A$ and $m_B$ but with the same velocity
(upon adopting (\ref{velPARADO}) and choosing appropriately
their different energies)
 according to some observer $O$.
So the starting point of the analysis
is such that the observer $O$ could see two particles with different masses $m_A$ and $m_B$
moving at the
same speed and following the same trajectory (for $O$ particles $A$ and $B$ are ``near"
at all times).
It is then interesting to investigate
how these two identical worldlines appear to a boosted observer $O'$.
And one easily finds~\cite{dsrIJMPrev}, using (\ref{velPARADO}) and (\ref{dsr1boostPARADO}),
that the two particles would have different velocities according
to a  second observer $O'$ (boosted with respect to $O$).
So according to $O'$ the two particles  could
be ``near" only for a limited amount of time.
This observation establishes the need for a nonclassical spacetime, since in particular
it implies that single points for $O$ are mapped into pairs of (possibly sizeably distant)
points for $O'$.

Similar conclusions are drawn by considering the implications of
(\ref{velPARADO}) and (\ref{dsr1boostPARADO}) for the analysis
of ``Einstein light clocks", the prototypical relativistic clocks,
which are
 digital clocks counting
the number of times a photon travels a known distance $L$ between two ideal mirrors.
Clearly in presence of energy dependence of the speed of photons, $v(E)$,
such a clock measures time in units of $\tau = v(E) L$,
and it is then straightforward to analyze the dilatation
of this time interval $\tau$ under a
boost acting in the direction orthogonal to the axis that connects
the two mirrors~\cite{dsr1,dsr2}:
\begin{equation}
\tau' = {v(E) \over \sqrt{v(E')^2 - V^2}} \ \tau ~ \label{tautau}
\end{equation}
where $V$ is the relative velocity of the reference frames connected
by the boost, $E$ is the energy of the photon in the rest frame of the clock,
and $E'$ is the energy of the photon in the boosted frame.

At first sight (\ref{tautau}) appears to be describable as an energy-dependent
but smooth deformation of the standard special-relativistic time-dilatation
formula, but upon closer examination one quickly realizes that it has strikingly
wild implications.
For my purposes here it suffices to take into account this
time-dilatation formula (\ref{tautau})
in the analysis of a pair of Einstein light clocks, one using photons
of energy $E_1$
and the other one, at rest with respect to the first,
using photons of energy $E_2$.
For simplicity, let me further specify that $E_1$ and $E_2$ be such that $v(E_2) = 2 v(E_1)$,
{\it i.e.} the difference in energy
is large enough to induce doubling of speed.
In such a setup the two clocks could in principle share the same pair
of mirrors, so that they provide a case with ``clocks at the same point",
which is of particular interest from a relativistic perspective.

This setup with two Einstein clocks is such that,
by construction/synchronization, in the rest frame of the clocks
one always finds
simultaneous "tick" of the $E_2$-clock whenever the $E_1$-clock "ticks",
since we have arranged things in such a way that
$$\tau_1 = 2 \tau_2~$$
And it is particularly significant that, using (\ref{dsr1boostPARADO}) to relate $E_1',E_2'$
to $E_1,E_2$, one finds
\begin{equation}
\tau_1' = {v(E_1) \over \sqrt{v(E_1')^2 - V^2}} \, \tau_1 \neq 2
\tau_2' = {v(E_2) \over \sqrt{v(E_2')^2 - V^2}} \, 2 \tau_2 ~
\label{tautauTWOCLOCKS}
\end{equation}
for the same boost considered above (rapidity $V$, orthogonal to the
axis of the mirrors). This implies that, while in their rest frame
the ticks of the $E_1$-clock are always simultaneous to ticks of the
$E_2$-clock, in the boosted frame the ticks of the $E_1'$-clock are
not simultaneous to ticks of the $E_2'$ clock. Ordinary special
relativity (while removing the abstraction of an ``absolute time")
still affords us an objective, observer independent, concept of
simultaneity, restricted to events occurring at the same spatial
point, but in this candidate DSR framework there are events ({\it
e.g.,} ticks of our two Einstein clocks) which are simultaneous and
at the same spatial point for observer $O$ but are not simultaneous
for observer $O'$. Using tentatively classical-spacetime reasoning
one finds that an ``event", a point of spacetime, for observer $O$
is not mapped in an event for $O'$, which in turn implies that the
classical-spacetime description of an event (as a sharp point in
spacetime) must be abandoned.

\subsection{Spacetime Fuzziness for Quantum Particles}
The previous subsection exposes the need for spacetime nonclassicality
already for (formal) theories of classical particles.
Let me now comment (following the line of
reasoning adopted in Ref.~\cite{bignapapPRD})
on the even more virulent departures from
spacetime classicality that are to be expected when a DSR framework predicts
energy dependence of the speed of {\underline{quantum}} photons.

The issue arises because in giving operative meaning to spacetime points
(or, more rigorously, to distances between spacetime points)
one inevitably resorts to the use of particle probes and spacetime
can be meaningfully labelled ``classical"
only if the theory admits
the possibility for such localization procedures to have
absolutely sharp (``no uncertainty") outcome,
at least as the endpoint of a well-defined limiting procedure.
Within ordinary quantum mechanics one could still legitimately contemplate
the ideal limit in which point particles have infinite inertial mass
(so that the
Heisenberg principle is ineffective) but that limit is not meaningful in
theoretical frameworks, such as the DSR framework, in which the motivation
originates from the quantum-gravity problem and therefore infinite inertial
mass comes along with infinite gravitational charge (mass).
In the opposite limit, the one of massless particles, one finds that the
energy(/wavelength) dependence of the speed of light introduces an
extra term in the balance of quantum uncertainties.
Assume in fact that the measurement procedure requires
some known time $T_{obs}$ and therefore (in order to obtain measurement results
compatible with the classical-spacetime idealization)
we would like the particle probe
to behave as a classical probe over that time.
For that goal it is in particular necessary to keep under control
two sources of quantum uncertainty, the one concerning the energy of
the particle and the one concerning the time of emission of the particle.
The uncertainty $\delta x$ in the position of the massless
probe when a time $T_{obs}$ has lapsed since the observer
(experimentalist) set off the measurement procedure will in general
satisfy the following inequality
\begin{equation}
\delta x \ge  \, \delta t + \delta v \, T_{obs} \simeq \delta t +
\lambda \delta E T_{obs} ~ \label{deltagacgen}
\end{equation}
where $\delta t$ is the uncertainty
in the time of emission of the probe,
and I used (\ref{velPARADO})
to  describe the uncertainty in the speed of the particle, $\delta v$,
in terms of the uncertainty $\delta E$
in its energy.
Since the uncertainty
in the time of emission of a particle and the uncertainty
in its energy are related\footnote{It is well understood
that the $\delta t \, \delta E \ge 1$ relation
is valid only in a weaker sense than, say,
Heisenberg's Uncertainty Principle $\delta x \, \delta p \ge 1$.
This has its roots in the fact  there is no self-adjoint operator canonically conjugate
to the total energy, if the energy spectrum is bounded
from below.
However, $\delta t \, \delta E \ge 1$
does relate $\delta t$ intended as uncertainty
in the time of emission of a particle and $\delta E$
intended as uncertainty
in the energy of that same particle, and therefore it applies in the context
which I am here considering.}
by $\delta t \, \delta E \ge 1$, Eq.~(\ref{deltagacgen})
can be turned into an {\underline{absolute bound}} on
the uncertainty in the position of the massless
probe when a time $T_{obs}$ has lapsed since the observer
set off the measurement procedure:
\begin{equation}
\delta x \ge  {1 \over \delta E} +\lambda \delta E  T_{obs} \ge
\sqrt{{\lambda T_{obs}}} ~ \label{deltagacgenfin}
\end{equation}
The right-hand side of (\ref{deltagacgenfin}) does exploit
the fact that in principle the observer
can prepare the probe in a state with desired $\delta E$
(so it is legitimate to minimize the uncertainty with respect
to the free choice of $\delta E$), but the classical behaviour
of the probe is not achieved in any case (in all cases $\delta x$
is {\underline{strictly greater}} than 0).

%JOC%%%%%%%%%%%%%%%%%
%
% a questa subsec su nonclassical spacetime devo aggiungere l'argomento
% su massive particles of same speed for O but different speed for O'
%
%

\section{More on the Use of Hopf Algebras in DSR Research}
I have already mentioned Hopf algebras as a candidate tool for the formalization
of DSR relativistic theories. In this section I briefly summarize the technical reasons
why this appears to be plausible, but also describe some technical challenges
which still need to be addressed before drawing any conclusions on the applicability
(or lack thereof)
of Hopf algebras to DSR model building. Also on this topic of the relevance of Hopf algebras
for DSR research some misconceptions are rather frequent, so I will find opportunities
for additional true/false characterizations.

\subsection{A Hopf-algebra Scenario with $\kappa$-Poincar\'e Etructure}
In trying to give a brief summary of some key aspects of the possibility
of DSR scenarios based on the mathematics of Hopf algebras, let me start with
a rudimentary review of the most studied such scenario which is based on
the $\kappa$-Poincar\'e Hopf algebra and
the associated $\kappa$-Minkowski noncommutative spacetime.

The characteristic spacetime-coordinate noncommutativity of $\kappa$-Minkowski is
given by
\begin{eqnarray}
&[x_j,x_0]= i \lambda x_j \\
&[x_k,x_j]=0 ~\label{kmnoncomm}
\end{eqnarray}
where $x_0$ is the time coordinate, $x_j$ are space coordinates ($j,k \in \{1,2,3\}$),
and $\lambda$ is a length scale, usually expected to be of the order of the
Planck length. Functions of these noncommuting coordinates
are usually
conventionally taken to be of the form
\begin{equation}
f(x) = \int d^4 k \tilde{f} (k) e^{i \vec{k} \cdot \vec{x}} e^{- i
k_0 x_0} \label{kappafourier}
\end{equation}
where the ``Fourier parameters" $\{k_0 , k_i \}$ are ordinary commutative variables.

I shall here be satisfied with a brief review of a frequently used
characterization of symmetries of $\kappa$-Minkowski, in which
generators for translations, space-rotations and boosts are introduced adopting
the following definitions
\begin{equation}
P_\mu \left( e^{i \vec{k} \cdot \vec{x}} e^{-ik_0x_0} \right) =
k_\mu e^{i \vec{k} \cdot \vec{x}} e^{-ik_0x_0} ~ \label{pmuaction}
\end{equation}
\begin{equation}
R_j \left( e^{i \vec{k} \cdot \vec{x}} e^{-ik_0x_0} \right) =
\epsilon_{jkl} x_k k_l e^{i \vec{k} \cdot \vec{x}} e^{-ik_0x_0}~
\end{equation}
\begin{equation}
N_j \left( e^{i \vec{k} \cdot \vec{x}} e^{-ik_0x_0} \right) = - k_j
e^{i \vec{k} \cdot \vec{x}} e^{-ik_0x_0} x_0 + \left[ x_j \left(
\frac{1 - e^{- 2 \lambda k_0}}{2 \lambda} + \frac{\lambda}{2}
|\vec{k}|^2\right) - \lambda x_l k_l k_j \right] e^{i \vec{k} \cdot
\vec{x}} e^{-ik_0x_0} ~
\end{equation}
The fact that we are here dealing with a Hopf algebra
(indeed the $\kappa$-Poincar\'e Hopf algebra)
is essentially seen by acting with these generators on products of functions
 (``coproduct"), observing for example that
\begin{eqnarray}
P_\mu \left( e^{i \vec{k} \cdot \vec{x}} e^{-ik_0x_0}
e^{i \vec{q} \cdot \vec{x}} e^{-iq_0x_0} \right)
&=& \left( k_\mu + e^{-\lambda k_0 (1-\delta_{\mu_0})} q_\mu \right)
\left( e^{i \vec{k} \cdot \vec{x}} e^{-ik_0x_0} e^{i \vec{q} \cdot \vec{x}} e^{-i q_0x_0} \right)
\nonumber\\
&\neq&\left( k_\mu + q_\mu \right) \left( e^{i \vec{k} \cdot
\vec{x}} e^{-ik_0x_0} e^{i \vec{q} \cdot \vec{x}} e^{-i q_0x_0}
\right) ~ \label{coprodpmu}
\end{eqnarray}
Some authors naively focus their analyses exclusively on
the properties of commutators of the generators
of the $\kappa$-Poincar\'e algebra, which turn out to be
deformations~\cite{aadluna} of
the commutators of the standard Poincar\'e Lie algebra.
And several (mutually incompatible) naive arguments
have been proposed on the basis of such naive characterizations: in some cases
the nonlinearities present in these commutators are taken as a full characterization
of the $\kappa$-Poincar\'e symmetries,
while in other cases a large significance is incorrectly given to
the fact that by
nonlinear redefinition of the generators
one can remove all anomalies of the commutators.
However, a proper description of Hopf-algebra spacetime symmetries
must clearly take into account both the commutators and the coproducts
(and redefinitions
of the generators affect simultaneously both commutators and coproducts).

Consider for example the case of translation transformations.
A translation transformation for a function $f(x)$ should be described as
\begin{equation}
df(x)=i\epsilon^\mu P_\mu f(x) \label{differenzialeP}
\end{equation}
in terms of the translation generators and some
transformation parameters $\epsilon_\mu$. And it turns out that the transformation
parameters must reflect the properties of the coproduct.
In fact, the transformation parameters must ensure that (if  $x_\mu$
is in $\kappa$-Minkowski) $x_\mu + \epsilon_\mu$
is still a point in $\kappa$-Minkowski:
\begin{eqnarray}
[x_j +\epsilon_{j},x_0+ \epsilon_0]= i \lambda
(x_j+\epsilon_{j})~,~~ [x_i+\epsilon_{i},x_j +\epsilon_{j}]=0
~\label{check}
\end{eqnarray}
So the transformation parameters must not be simple numbers but should instead
be endowed with nontrivial algebraic properties. And it is easy
to see~\cite{kappanoether,nopureboost} that the form of these algebraic properties
should reflect the properties of the coproduct in order to preserve Leibniz rule:
\begin{equation}
d(f\cdot g)=f\cdot dg+df \cdot g ~\label{leib}
\end{equation}
Taking into account these observations it turned out to be
possible~\cite{kappanoether,nopureboost}
to obtain conserved charges associated to the Hopf symmetries
for a theory with classical fields in the noncommutative $\kappa$-Minkowski spacetime
(while all previous attempts, which had naively ignored the role of the
coproduct in the full characterization of symmetry transformations, had failed).

Besides providing insight on the role on nonlinear redefinitions of generators
within Hopf-algebra formulations (already here stressed in Subsection~3.1),
these results on ``noncommutative transformation parameters" might have
other strong implications. As first stressed in Ref.~\cite{nopureboost},
by examining in detail the nonocmmutativity properties of the transformation
parameters one finds that pure boosts are forbidden: whenever
the boost parameters are nonzero the noncommutativity properties are such that
also some rotation parameters should necessarily be turned on~\cite{nopureboost}.
This is rather significant for DSR scenarios which have been tentatively based
on the $\kappa$-Poincar\'e/$\kappa$-Minkowski: it seems we should assume that,
if eventually we do find a fully consistent DSR scenario based on
the  $\kappa$-Poincar\'e/$\kappa$-Minkowski formalism then
the description of boosts should accommodate some nonclassical features
even in the characterization of the differences between pairs of observers.

But let me stop here my brief summary of results obtained in
the $\kappa$-Poincar\'e/$\kappa$-Minkowski framework,
since it will suffice for my purposes.
The interested reader can find a rather detailed description of what we presently
know about the $\kappa$-Poincar\'e symmetries of $\kappa$-Minkowski in
Refs.~\cite{kappanoether,nopureboost} and references therein.
The idea that this mathematics might provide the basis for a DSR theory
originates essentially in the observation that the spacetime noncommutativity
of $\kappa$-Minkowski, as described in (\ref{kmnoncomm}),
is left invariant by the action of $\kappa$-Poincar\'e generators,
and the scale $\lambda$ appears to be a reasonable candidate
for a DSR-type second relativistic scale.
It has also long been conjectured that with
this $\kappa$-Minkowski/$\kappa$-Poincar\'e recipe one might end up having
a DSR theory with modified dispersion relation, but originally this suggestion
was only based on the observation that the ``mass Casimir" of
the $\kappa$-Poincar\'e Hopf algebra is a deformation of the mass Casimir of the
standard Poincar\'e Lie algebra. A definite statement about the status
of the dispersion relation in this framework will require a meaningfully
physical identification and characterization of concepts such as energy,
spatial momentum, frequency and wavelength, and this task is perhaps
now finally
within reach, since we can now rely on actual derivations of conserved charges
using the techniques developed in Refs.~\cite{kappanoether,nopureboost}. However,
a few residual issues must still  be
addressed~\cite{kappanoether,nopureboost,kowa5d,kappanoether5D}
before we can safely identify energy,
spatial momentum, frequency and wavelength.

And in general, as mentioned in earlier sections of these notes, some
work remains to be done to fully establish
that the $\kappa$-Minkowski/$\kappa$-Poincar\'e formalism
can be really used to construct
a DSR theory. It will require still some work
of digging through $\kappa$-Minkowski/$\kappa$-Poincar\'e mathematics,
guided by the DSR
principles, looking for tools that are DSR compatible.
But this is not trivial and not easily done without caution. For example,
as mentioned,
the $\kappa$-Poincar\'e mathematics can inspire (and has inspired) a description
of the kinematics of particle-reaction processes which is manifestly
in conflict with the DSR principles (it would select a preferred frame),
but looking around in the $\kappa$-Minkowski/$\kappa$-Poincar\'e ``zoo of mathematics",
if we indeed look around using the DSR
principles as guidance, we might find other (possibly DSR-compatible)
structures on which
the kinematics of particle-reaction processes could be based.

\subsection{A Hopf-algebra Scenario without  $\kappa$-Poincar\'e
and without Modified Dispersion Relations} The fact that the
$\kappa$-Poincar\'e/$\kappa$-Minkowski framework briefly discussed
in the previous subsection provides a promising path toward a DSR
theory is appreciated by most authors involved in DSR research. So
much so that some do not appear to even see the possibility of
alternatives. Actually, one does not need to look too far to find an
alternative which is equally promising: even within the confines of
approaches based on spacetime noncommutativity one finds a framework
which is indeed equally promising from a DSR perspective and appears
to motivate investigation of a different set of candidate ``DSR
effects". This is the case of the ``canonical noncommutative
spacetime" with characteristic spacetime-coordinate noncommutativity
given by ($\mu,\nu \in \{0,1,2,3\}$)
\begin{equation}
[{x}^\mu,{x}^\nu]=i\theta^{\mu\nu} ~ \label{thetacr}
\end{equation}
These noncommutativity relations (\ref{thetacr}) can be meaningfully considered
endowing $\theta^{\mu\nu}$ with the properties of a standard Lorentz tensor,
in which case one of course ends up with a scenario where Lorentz symmetry is
broken, but they can also be meaningfully considered
endowing $\theta^{\mu\nu}$ with the properties of an observer-independent
matrix, which would of course be the case of interest from a DSR perspective.
[A third, even more ambitious, possibility has been developed in Refs.~\cite{doplich1,doplich2}
by allowing $\theta^{\mu\nu}$ to acquire nontrivial algebraic properties.]

With observer-independent $\theta^{\mu\nu}$
the symmetries of this spacetime are described by a Hopf algebra
which is significantly {\underline{different from $\kappa$-Poincar\'e}},
whose generators can be described as follows:
\begin{eqnarray}
&P_\mu e^{ik x}&\equiv P_\mu \Omega (e^{ikx^{(c)}})
 \equiv \Omega(i\partial_\mu^{(c)} e^{ikx^{(c)}}) \label{classicalWa} \\
&M_{\mu\nu} e^{ik  x}&\equiv M_{\mu\nu} \Omega (e^{ikx^{(c)}})\equiv
\Omega(ix^{(c)}_{[\mu}\partial^{(c)}_{\nu]} e^{ikx^{(c)}}) ~
\label{classicalWb}
\end{eqnarray}
where $x^{(c)}_{\mu}$ are auxiliary commutative coordinates,
 $\partial^{(c)}_{\nu}$ are ordinary derivatives with respect to
 the auxiliary coordinates $x^{(c)}_{\mu}$,
and (for any $\tilde\Phi (k)$)
\begin{equation}
\Omega \left( \int d^4 k\, \tilde\Phi (k) e^{ikx^{(c)}} \right)=\int
d^4k \,\tilde\Phi (k) e^{ik x} ~ \label{weylmap}
\end{equation}

The fact that these are generators of a Hopf algebra
manifests itself as usual immediately upon noticing
that the
action of (Lorentz-sector) generators does not comply with Leibniz
rule:
\begin{eqnarray}
M_{\mu \nu}\left(e^{ik x}e^{i q x}\right)&=& \left(M_{\mu \nu}
e^{ik x}\right)e^{iq x}+  e^{ik x} \left(M_{\mu \nu} e^{iq x}\right)+\nonumber\\
&& + \frac{1}{2}\theta^{\alpha\beta}\left[- \eta_{\alpha
[\mu}\left(P_{\nu]} e^{ik x}\right)\left( P_\beta e^{iq x} \right) -
\left(P_\alpha e^{ik x}\right) \eta_{\beta [\mu} \left(P_{\nu]}
e^{iq x}\right)\right] ~\label{actiononexponentialsM}
\end{eqnarray}

As in the case of the $\kappa$-Poincar\'e/$\kappa$-Minkowski
framework, the idea that this framework based on canonical
noncommutativity might provide the basis for a DSR theory originates
essentially in the observation that the spacetime noncommutativity
of canonical form, (\ref{thetacr}), is left invariant by the action
of the Hopf-algebra generators
(\ref{classicalWa})-(\ref{classicalWb}), so that any physical
consequence of that noncommutativity (such as ``spacetime
fuzzyness") should be observer independent. To render explicit the
presence of a short invariant length scale one can conveniently
rewrite the observer-independent dimensionful matrix
$\theta^{\mu\nu}$ in terms of an observer-independent length scale
$\lambda$ and an observer-independent dimensionless matrix
$\tau^{\mu\nu}$ (with an extra restriction, {\it e.g.,} unit
determinant, to avoid apparent overcounting of parameters upon
introducing $\lambda$): $\theta^{\mu\nu} = \lambda^2 \tau^{\mu\nu}$.
And the length scale $\lambda$ should indeed be small in order to
ensure that one introduces an acceptably small amount of coordinate
noncommutativity (we clearly have robust experimental evidence
against large noncommutativity).

For a detailed description of what we presently
know about this alternative Hopf-algebra-based DSR scenario
readers can look at Ref.~\cite{thetanoether,nopureboost} and references therein.
One key point is that all evidence gathered so far (which however only concerns
classical particles and fields in this ``quantum" geometry)
suggests that the dispersion relation is not modified in this framework,
and therefore also from this perspective we might already have a framework
providing complementary DSR intuition with respect to
the more popular $\kappa$-Poincar\'e/$\kappa$-Minkowski
framework.
Also for this canonical-noncommutativity framework it turned out
to be possible~\cite{thetanoether}
to derive conserved charges within an analysis {\it a la}
Noether, and in doing this it turned out to be necessary
to endow the transformation parameters with nontrivial algebraic
properties reflecting the coproduct structure.
And the  choice of ordering implicitly introduced in (\ref{weylmap})
in this case can be shown very explicitly  to be inessential:
a variety of alternative ordering conventions are easily considered
and found~\cite{thetanoether} to lead to the same conserved charges.
Since by changing ordering convention for the spacetime coordinates
one essentially ends up introducing nonlinear redefinitions
of the generators of the Hopf algebra, the fact that
the charges are independent of the choice of ordering convention
also translates into yet another invitation for caution for those authors
who naively assume that nonlinear redefinitions of the generators
of a Hopf algebra might change the physical picture (in some extremely
naive arguments it is stated that by nonlinear redefinition of the generators
one might go from an incorrect to a correct picture, or that such a nonlinear
redefinition of generators could eliminate the characteristic length scale
of some relevant Hopf algebra, but here very explicitly one finds that such nonlinear
redefinitions do not affect the conserved charges and their dependence
on the characteristic scale of the Hopf algebra).

\section{DSR Scenarios and DSR Phenomenology}
Considering the very early stage of development of the Hopf-algebra
scenarios for DSR, and the fact that the approach based on Hopf algebra
is the best developed
%JOC%(the least underdeveloped)
attempt of finding DSR compatible theories,
it is clear that we are not ready to do any ``real" DSR phenomenology.
In order to claim one was doing real DSR phenomenology a minimum requisite
would be the availability of a theoretical framework whose compatibility
with the DSR principles was fully established and characterized in
terms of genuinely observable features.
Since we are not ready for that, one might perhaps consider postponing
all reasoning about phenomenology to better times (ahead?).
But, at least from some perspectives the development of
(however incomplete and however {\it ad hoc}) ``toy DSR test theories"
can be valuable. The exercise of developing such ``toy test theories",
in the sense that will emerge from the next subsections,
turns out to be valuable in providing a crisper physical characterization
of the concept of a DSR theory (and in fact I found it useful
to introduce one such ``toy theory" already in my original papers in
Refs.~\cite{dsr1,dsr2}) and allows to clarify some general arguments
(as in the case of the incompatibility of a decay threshold with the
DSR principles, see later).

\subsection{A Toy-Model DSR-Scenario Test Theory, Confined to Leading Order}
To illustrate what I qualify as a ``toy DSR test theory"
I can indeed use the example of
such ``toy theory" that I already used in my original papers in
Refs.~\cite{dsr1,dsr2}.
This is a ``limited theory" in that it only concerns the laws of transformation
of the energy-momentum observables\footnote{Of course, an analogous ``limited theory"
(with the same formulas) could be articulated for the frequency/wavelength
observables and for laws governing the composition and onshellness
of frequencies/wavelengths.}..
It assumes that the energy-momentum dispersion relation is observer independent
and takes the following form in leading order in the Planck length:
\begin{equation}
 E^2 \simeq  \vec{p}^2 + m^2 + \lambda \vec{p}^2 E
  ~\label{dispKpoinnew}
\end{equation}
This dispersion relation is clearly an invariant of space
rotations, but it is not an invariant of ordinary boost
transformations.
Its invariance (to leading order) is ensured adopting  standard
space-rotation generators
\begin{equation}
R_j = - i \epsilon_{jkl} p_k {\partial \over \partial p_l} ~
\label{rotnormal}
\end{equation}
and a deformed action for boost generators
\begin{equation}
{\cal B}_j \simeq i p_j \frac{\partial}{\partial E}+ i \left(E +
\frac{\lambda}{2} {\vec p}^2 + \lambda E^2 \right)
\frac{\partial}{\partial p_j} -i \lambda p_j  \left(p_k
\frac{\partial}{\partial p_k} \right )~ \label{dsr1boosts}
\end{equation}

For the rudimentary ``phenomenology of kinematics" which I intend to discuss
within this limited  ``toy DSR test theory"
the only remaining ingredient to be specified is the one linking
incoming energymomenta to outgoing ones, as intended in a law
of  conservation of energymomentum.
Let us start considering processes  with
two incoming particles, $a$ and $b$, and two outgoing
particles, $c$ and $d$.
The special-relativistic kinematic requirements for
such processes are $E_a + E_b - E_c -E_d=0$ and $p_a + p_b - p_c -p_d=0$,
but these clearly~\cite{dsr1,dsr2} would not be observer-independent laws
in light of (\ref{dsr1boosts}).
Working in leading order actually one finds several~\cite{dsr1,dsr2}
acceptable\footnote{The conservation laws
that must be satisfied by physical processes
should~\cite{dsr1dsr2} be covariant
under the transformations that relate the kinematic properties
of particles as measured by different observers
(all observers should agree on whether or not
a certain process is allowed).}
alternative possibilities for the
deformation of the law of conservation of energymomentum.
In the following I will adopt
\begin{equation}
E_a + E_b + \lambda p_a p_b \simeq
 E_c +E_d +\lambda p_c p_d
~ \label{conservnewe}
\end{equation}
\begin{equation}
p_a + p_b + \lambda (E_a p_b + E_b p_a) \simeq
 p_c +p_d + \lambda (E_c p_d + E_d p_c)
~ \label{conservnewp}
\end{equation}

Analogous formulas can be obtained for any process with $n$
 incoming particles and $m$ outgoing
particles.
In particular, in the case of a
two-body particle decay $a \rightarrow b+c$
the laws
\begin{equation}
E_a \simeq E_b + E_c + \lambda {p}_b  {p}_c ~ \label{consDecayA}
\end{equation}
\begin{equation}
p_a \simeq p_b + p_c + \lambda (E_b p_c + E_c p_b) ~
\label{consDecayB}
\end{equation}
provide an acceptable (observer-independent, covariant according to
(\ref{dsr1boosts})) possibility.

\subsection{On the Test Theory Viewed from An (unnecessary) All-Order Perspective}
As stressed earlier in these notes, the physical picture that
motivates the proposal of Doubly-special Relativity allows one to
contemplate DSR symmetries as exact symmetries (applicable in a
certain corresponding regime), but it also invites one to consider
the possibility that (even within the confines of the regime where
DSR turns out to be applicable) DSR symmetries be just approximate
symmetries. For example, formulas might not be exactly compatible
with the DSR setup because of possibly including non-DSR terms that
are negligible (but nonzero) in the DSR regime, but become large at
some even higher energy scales, where quantum-gravity effects might
become so virulent not to allow even a \linebreak\ DSR description.

I should therefore correspondingly stress that it is not necessary
(and not necessarily appropriate) to cast the ``leading-order toy test theory"
discussed in the previous subsection within some corresponding all-order
DSR theory.

Let me nonetheless briefly review some evidence we have that there is
an ``all-order toy test theory" from which the leading-order one discussed
in the previous subsection can be derived.
A first ingredient for such a theory could be the following all-order
dispersion relation
\begin{equation}
{2 \over \lambda^2} \left[ \cosh ({\lambda E }) - \cosh ({\lambda
m}) \right] = \vec{p}^2 e^{\lambda E} ~\label{dispKpoinnewAO}
\end{equation}
in which case boost transformations
could be generated by
\begin{equation}
{\cal B}_j = i p_j \frac{\partial}{\partial E}+ i
\left(\frac{\lambda}{2} {\vec p}^2 - \frac{1-e^{2 \lambda E}}{2\lambda}\right
) \frac{\partial}{\partial p_j}-i \lambda p_j \left(p_k
\frac{\partial}{\partial p_k} \right )~ \label{dsr1boostsexa}
\end{equation}
(whereas, even in the all-order formulation,
space-rotation transformations do not require deformation).

Even for these ``all-order boost generators" (and therefore, of course,
also for their leading-order formulation)
one manages to obtain explicit formulas~\cite{gacRoxJurek} for the finite boost
transformations that relate the observations of two observers.
These are obtained by integrating the familiar differential
equations
\begin{equation}
{dE \over d\xi} = i [{\cal B}_j,E] ~,~~~{dp_k \over d\xi} = i [{\cal
B}_j,p_k] ~ \label{infp}
\end{equation}
which relate the variations of energy-moomentum with rapidity
($\xi$) to the commutators between the boost generator (along the direction $j$ of
the chosen boost) and
energy-momentum.

The result is conveniently characterized by giving the formula
expressing the amount of rapidity $\xi$ needed to take a particle from its
rest frame (where the energy is $m$)
to a frame in which its energy is $E$:
\begin{equation}\label{xidsr1new}
\cosh (\xi) = \frac{e^{\lambda E} - \cosh\left(\lambda m\right)}
  {\sinh\left( \lambda m \right)} ~,~~~
\sinh (\xi) = \frac{p e^{\lambda E}}
  {\lambda \sinh\left(\lambda m\right)}\,\,
\end{equation}
Since I am here only setting up a discussion of DSR phenomenology
I shall not make any further effort of
characterization of this all-order setup.
In fact, at least for teh foreseeable future, it appears likely
that we
could only attempt a leading order
DSR phenomenology (sensitivities presently available do not appear to
allow us to go beyond that).
My discussion of an all-order DSR setup clearly was not aiming for an ``all-order DSR
phenomenology", but rather to stress that if one insists (and it is not necessary but
possible to insist)  on embedding the ``leading order toy test theory"
within an all-order setup, then a valuable constraint can emerge:
while within the leading-order toy test theory
nothing prevents the (dimensionful) parameter $\lambda$ to be either
positive or negative, within the specific all-order
reformulation I just discussed $\lambda$ must be positive\footnote{Meaning that $\lambda/L_p$
is a positive real number.}.
This comes about~\cite{dsrphen} because attempting to find real
energy-momentum solutions
for Eqs.~(\ref{infp}) for all real values of $\xi$ (which one ought to do
in an all-order setup) and using the form (\ref{dsr1boostsexa}) of the boost
generators one actually finds the solutions codified in (\ref{xidsr1new})
only for positive $\lambda$, while the same equations for negative $\lambda$
do not admit solution (not solutions for all values of $\xi$ if the energy-momentum
must be real~\cite{dsrphen}).
Note however that the embedding of a ``leading order toy test theory"
within an ``all-order toy test theory"
is clearly not unique, and there may well be other such embeddings which instead
are admissible only for negative $\lambda$. One can therefore tentatively conclude that
 we should not expect a definite
sign prediction generically for the large class of possible DSR scenarios of this sort,
but that within any given all-order formulation, with of course some associated leading-order formulation,
 the sign ambiguity can be studied and
possibly removed (as it is removed in the specific all-order scenario discussed in this
subsection).

\subsection{Photon Stability}
The first possibility that I want to consider in this phenomenology
section is the one of energy thresholds for particle decay, such as
the possibility of the decay of a photon into an electron-positron
pair,  $\gamma \rightarrow e^+ + e^-$. I already discussed this
process earlier in these notes, as a way to show that the DSR
concept makes definite predictions, in spite of not being at present
attached to any specific mathematical formalism. Any experiment
providing evidence of a preferred frame would rule out DSR, and this
implies that any theory compatible with the DSR principle must not
predict energy thresholds for the decay of particles. In fact, one
could not state observer-independently a law setting a threshold
energy for a certain particle decay, because different observers
attribute different energy to a particle (so then the particle
should be decaying according to some observers while being stable
according to \linebreak\ other observers).

This argument holds directly at the level of the logical structure
of the DSR concept, but it is nonetheless useful to verify how our
toy DSR test theory implements it.
The key structure is the rigidity that the DSR concept introduces (for theories
structured like our toy test theory) between the structure of the energy-momentum
dispersion relation and the structure of the energy-momentum-conservation law.
The boost transformations must leave invariant the dispersion relation, and under those
same boost transformations the energy-momentum-conservation law must be covariant.
This led me in particular to introduce the
 energy-momentum-conservation law
\begin{equation}
E_a \simeq E_b + E_c + \lambda {p}_b  {p}_c ~ \label{consDecayAn}
\end{equation}
\begin{equation}
p_a = p_b + p_c + \lambda (E_b p_c + E_c p_b) ~ \label{consDecayBn}
\end{equation}
for two-body particle decays $a \rightarrow b+c$,
in association with the dispersion relation
\begin{equation}
 E^2 \simeq  \vec{p}^2 + m^2 + \lambda \vec{p}^2 E
  ~\label{dispKpoinnewnstab}
\end{equation}

If one was to combine (\ref{dispKpoinnewnstab}) with an unmodified law of
energy-momentum conservation, as admissible in scenarios with broken Poincar\'e
symmetry (but not allowed in scenarios in which Poincar\'e symmetry
is deformed in the DSR sense),
then it is well established
that a threshold for the decay of a photon into electron-positron pairs
can emerge. In the symmetry-breaking case the relation between
the energy $E_\gamma$ of the incoming photon, the opening angle
$\theta$ between the outgoing electron-positron pair, and the
energy $E_+$ of the outgoing positron,  takes the form
$\cos(\theta) = (A+B)/A$, where (for the region of
phase space with $m_e \ll E_\gamma \ll E_p$)
$A=E_+ (E_\gamma -E_+)$
and $B=  m_e^2 +
\lambda  E_\gamma E_+ (E_\gamma -E_+)$
 ($m_e$ denotes of course the electron mass), and  for $\lambda < 0$
 (meaning $\lambda/L_p$ negative real number) one finds
 that $\cos(\theta) < 1$ (and therefore the decay process is allowed)
 in certain corresponding region of phase space.

Instead if the same analysis is done in the DSR-compatible framework of
our toy DSR test theory, and therefore one adopts the modified dispersion
relation (\ref{dispKpoinnewnstab}) and the
modified energy-momentum-conservation
law (\ref{consDecayAn})-(\ref{consDecayBn}),
one arrives at a result for $\cos(\theta)$ which is still of the form $(A+B)/A$
but now with $A = 2 E_+ (E_\gamma -E_+) +
\lambda  E_\gamma E_+ (E_\gamma -E_+)$ and $B=2 m_e^2$.
Evidently this formula is never compatible with $\cos(\theta) < 1$,
consistently with the fact
that $\gamma \rightarrow e^+ e^-$ is always forbidden in our toy DSR
test theory (so, in particular, there is no threshold for the decay).

This discussion also shows at least one way in which toy DSR test theories
such as the one I am considering, in spite of all their limitations
for what concerns applicability
(the one I am considering only gives a rough description of some aspects
of kinematics)
and motivation (not being derived from, and not even inspired by,
a  DSR theory satisfactorily applied to a broad range of phenomena),
can be valuable in DSR research.
For example,  by analyzing decay processes within the framework
of a toy test theory
one can see explicitly the DSR principles at work in preventing
the emergence of particle-decay thresholds,
and this would be particularly valuable
if one had not noticed the general DSR argument
that forbids such decay thresholds.

\subsection{Weak Threshold Anomalies for Particle Reactions}
In the recent wide literature on a possible Planck-scale breakdown
of Lorentz symmetry there has been strong interest in the possibility
of large ``anomalies" in the evaluation of certain energy
thresholds for particle reactions that are relevant in
astrophysics.
A simple way to see this is found in the analysis
of collisions between
a soft photon of energy $\epsilon$
and a high-energy photon of energy $E$ that create an electron-positron
pair: $\gamma \gamma \rightarrow e^+ e^-$.
For given soft-photon energy $\epsilon$,
the process is allowed only if $E$ is greater than a certain
threshold energy $E_{th}$ which depends on $\epsilon$ and $m_e^2$.
In a broken-Lorentz-symmetry scenario
this threshold energy could be  evaluated combining
the dispersion relation  (\ref{dispKpoinnewnstab}) with ordinary
energy-momentum conservation, and this leads to the
result (assuming $\epsilon \ll m_e \ll E_{th} \ll 1/\lambda$)
\begin{equation}
E_{th} \epsilon + \lambda \frac{E_{th}^3}{8}= m_e^2 ~ \label{thrTRE}
\end{equation}
The special-relativistic result $E_{th} = m_e^2 /\epsilon$
corresponds of course to the $\lambda \rightarrow 0$ limit of
(\ref{thrTRE}). The Planck-scale ($\lambda$) correction can be
safely neglected as long as $\epsilon /E_{th} > \lambda E_{th}$. But
eventually, for sufficiently small values of $\epsilon$ and
correspondingly large values of $E_{th}$, the Planck-scale
correction cannot be ignored. This occurs for $\epsilon < (\lambda
m_e^4)^{1/3}$ (when, correspondingly $E_{th} >
(m_e^2/\lambda)^{1/3}$), and can be relevant for the analysis of
observations of multi-$TeV$ photons from certain
Blazars~\cite{aus,gactp}. Performing a similar analysis for the
photo-pion-production process, in which one has a high-energy proton
and a soft photon as incoming particles and a proton and a pion as
outgoing particles, one ends up finding~\cite{kifune,ita,gactp} the
possibility of large effects for the analysis of the cosmic-ray
spectrum in the neighborhood of the \linebreak ``GZK scale".

One might wonder whether something similar to
these particle-reaction-threshold results that are
so popular in the Lorentz-breaking literature could be found
also in a DSR setup.
In the case of photon stability we managed to draw a robust conclusion
in spite of the present limited understanding and development
of DSR theories, and our toy DSR test theory provided an explicit illustration
of how the DSR principles intervene in the analysis.
In the case of the analysis of particle-reaction thresholds
the indications we can presently give from a DSR perspective
are not as strong, but still rather valuable.
The possibility of large particle-reaction threshold anomalies
%(even if they were as large as in some Lorentz-breaking scenario)
cannot be excluded simply on the basis of the DSR principles,
since it does not conflict with any aspect of the structure of those principles,
nonetheless, by trial and error, the DSR literature has provided evidence
that large threshold anomalies are not naturally accommodated
in a DSR framework.
A key point to understand is that typically a DSR framework
 will produce smaller anomalies than a typical
symmetry-breaking framework. This is to be expected simply because
scenarios with deformation of symmetries produce of course softer departures from
the original symmetries than scenarios in which those symmetries
are broken (deforming is softer than breaking).

It is difficult to convey faithfully here in a few pages how
the DSR literature provides support for this intuition,
but I can quickly discuss how our toy DSR test theory deals with
particle-reaction thresholds, and this will to some extent
allow readers to start forming their own intuition.
The laws (\ref{conservnewe})-(\ref{conservnewp}) for the case
of  $\gamma \gamma \rightarrow e^+ e^-$
take the form~\cite{dsr1,dsr2,sethdsr,dsrphen}
\begin{eqnarray}
E + \epsilon - \lambda \vec{P} {\cdot} \vec{p} \simeq E_+
+ E_- - \lambda \vec{p}_+ {\cdot} \vec{p}_- ~,~~~
 \vec{P} +  \vec{p}
 + \lambda E \vec{p}
 + \lambda \epsilon \vec{P}  \simeq \vec{p}_+ + \vec{p}_-
 + \lambda E_+ \vec{p}_-
 + \lambda E_- \vec{p}_+ ~
\label{consgammagamma}
%\end{equation}
\end{eqnarray}
where I denoted with $\vec{P}$ the momentum of the photon of energy $E$
and I denoted with  $\vec{p}$ the momentum of the photon
of energy $\epsilon$.

Using these (\ref{consgammagamma}) and the dispersion relation
of our toy test theory  (\ref{dispKpoinnewnstab}) one obtains
(keeping only terms that are meaningful
for $\epsilon \ll m_e \ll E_{th} \ll 1/\lambda$)
\begin{equation}
E_{th} \simeq \frac{m_e^2}{\epsilon } ~ \label{thrTREbis}
\end{equation}
{\it i.e.} one ends up
with the same result as in the special-relativistic case.

This indicates that there is no large threshold anomaly in our
toy DSR test theory.
Actually this test theory does predict a small threshold anomaly,
but truly much smaller than discussed in some symmetry-breaking scenarios.
If, rather than working in leading order within the approximations allowed
by the hierarchy $\epsilon \ll m_e \ll E_{th} \ll 1/\lambda$,
one derives a  DSR threshold formula
of more general validity within our toy DSR test theory, one does find
a result which is
different from the special-relativistic one, but the differences are quantitatively
much smaller than in some known symmetry-breaking frameworks (and in particular
in the DSR case, unlike some symmetry-breaking scenarios,
the differences are indeed negligible
for $\epsilon \ll m_e \ll E_{th} \ll 1/\lambda$).

By trial and error one ends up figuring out that the same outcome
is obtained in other toy DSR test theories, if they are all based
on dispersion relations roughly of the type of (\ref{dispKpoinnewnstab}).
Note however that by ad hoc choice of the dispersion relation one can modify
our toy test theory in such a way to produce a large threshold anomaly. This is
for example accomplished~\cite{dsrgzk} by adopting a dispersion relation
which leads to the following energy-rapidity relation~\cite{dsrgzk}
\begin{equation}\label{dispgoodOK}
\cosh (\xi) = {E \over m} (2 \pi)^{-E^2 \tanh[m^2/(\lambda^4
E^6)]/(E^2 +m/\lambda)} \,\, ~
\end{equation}
but it might be inappropriate to attach much significance to such an ad hoc
setup.

In closing this discussion of thresholds for particle-reaction thresholds
let me just observe that in a theory that preserves the equivalence
of inertial frames the threshold conditions must be written
as a comparison of invariants. For example,
it is no accident that in special relativity
the threshold condition for $\gamma \gamma \rightarrow e^+ e^-$
takes the form $E \epsilon \ge m_e^2$.
In fact, $m_e^2$ is of course a special-relativistic invariant
and $E \epsilon$ is also an invariant (for the head-on collision
of a photon with four momentum $P_\mu$, such that $P_0=E$,
and a photon with four momentum $p_\mu$, such that $p_0= \epsilon$,
one finds, also using the special-relativistic dispersion
relation, that $P_\mu p^\mu = 2 E \epsilon$).
The fact that in constructing a DSR framework, by definition, one must
also insist on the equivalence
of inertial frames, implies that in any genuine DSR framework
the threshold conditions must also be written
as a comparison of invariants.

\subsection{Wavelength Dependence of the Speed of Light}
The structure I have introduced so far in our toy test theory does
not suffice to derive an energy dependence\footnote{The title of
this subsection mentions a wavelength dependence while the
discussion considers an energy dependence. This may serve as a
reminder of the two possibilities, which should however be treated
as distinct possibilities~\cite{dsrIJMPrev}. One may or may not
have a DSR deformation of the energy-momentum relation and one may
or may not have a DSR deformation of the wavelength-frequency
relation. The two structures are linked by standard
energy-frequency and momentum-wavelength relations in our ordinary
(pre-quantum-gravity) theories, but those relations may also be
DSR deformed, and there are arguments~\cite{dsrIJMPrev} suggesting
that there might be a deformation of the wavelength-frequency relation
without an associated deformation of the energy-momentum relation
(a scenario which of course would require deformation of the
energy-frequency and momentum-wavelength relations).}
of the speed of
photons, but this of course will emerge if one assumes that the
standard formula $v=dE/dp$ applies.
Assuming $v=dE/dp$ our test theory leads to the following velocity
formula (for $m < E \ll E_p \sim 1/ \lambda$):
\begin{equation}
v \simeq 1 - \frac{m^2}{2 E^2} + \lambda E ~ \label{velLIVbis}
\end{equation}
and there is of course a rather direct way to investigate
the possibility (\ref{velLIVbis}):
whereas in ordinary special relativity two photons ($m=0$)
with different energies
emitted simultaneously would reach simultaneously a far-away detector,
those two photons should reach the detector at different times
according to (\ref{velLIVbis}).

This type of effect emerging from
an energy dependence of the speed of photons
can be significant~\cite{grbgac,biller}
in the analysis of short-duration gamma-ray bursts that reach
us from cosmological distances.
For a gamma-ray burst it is not uncommon
to find a time travelled
before reaching our Earth detectors of order $T \sim 10^{17} s$.
Microbursts within a burst can have very short duration,
as short as $10^{-3} s$ (or even $10^{-4} s$), and this means that the photons
that compose such a microburst are all emitted at the same time,
up to an uncertainty of $10^{-3} s$.
Some of the photons in these bursts
have energies that extend at least up to the $GeV$ range.
For two photons with energy difference of order $\Delta E \sim 1 GeV$
a $\lambda \Delta E$ speed difference over a time of travel
of $10^{17} s$
would lead to a difference in times of arrival of
order $\Delta t \sim \lambda T \Delta E/E_p \sim 10^{-2} s$, which
is significant (the time-of-arrival differences would be larger than
the time-of-emission differences within a single microburst).

So clearly observations of gamma-ray bursts should eventually provide
valuable tests of DSR scenarios of this sort. At present however we must proceed cautiously,
since all results so far reported in the DSR literature concern flat-space applications,
whereas evidently spacetime curvature plays an important role in the analysis of
gamma-ray bursts emitted from cosmologically distant sources.
Of course, the ultimate goal of DSR research should be ``DSR-compatible geometrodynamics"
(a description of gravitational phenomena compatible with the DSR principles),
but it seems that in order to profit from the opportunity that gamma-ray bursts provide
it would suffice to make the next step on the way from DSR in flat spacetime
to DSR geometrodynamics, {\it i.e.} the formulation of scenarios for the DSR principle
to be applied in a background (nondynamical) spacetime geometry with curvature.

\subsection{Crab-nebula Synchrotron Radiation Data}
For studies of scenarios in which Lorentz symmetry is broken by Planck-scale
effects another valuable opportunity is provided by observations
of the Crab nebula which are naturally interpreted as the result
of synchrotron-radiation emission.
This is part of the studies aimed at testing the possibility
of energy dependence of the speed of particles of the
type
\begin{equation}
v \simeq 1 - \frac{m^2}{2 E^2} + \lambda E ~ \label{velLIVcrab}
\end{equation}
within scenarios in which Lorentz symmetry is broken.
Assuming that all other aspects of the analysis of synchrotron
radiation remain unmodified at the Planck scale, one
is led~\cite{jaconature} to the conclusion that, if $\lambda$ is negative
($|\lambda| \sim L_p$,  $\lambda/L_p < 0$), the
energy dependence of the Planck-scale ($\lambda$) term in (\ref{velLIVcrab})
can severely affect the value of the cutoff energy for
synchrotron radiation~\cite{jackson}.
In fact, for negative $\lambda$, an electron on, say, a circular trajectory
(which therefore could emit synchrotron radiation)
cannot have a speed that exceeds the maximum
value
\begin{equation}
v^{max}_{e} \simeq 1 - \frac{3}{2} \left( |\lambda| m_e
\right)^{2/3} ~ \label{velLIVmax}
\end{equation}
whereas in special relativity of course $v^{max}_{e} = 1$ (although values
of $v_e$ that are close to $1$ require a very large electron energy).

This may be used to argue that for negative $\lambda$
the cutoff energy for synchrotron radiation
should be lower than it appears to be suggested by certain
observations of the Crab nebula~\cite{jaconature}.

Only very little of relevant for this phenomenology can be said
at present from a DSR perspective.
One key concern is that we have at present a very limited
understanding of what should typically characterize interactions in a DSR framework,
and a proper analysis of synchrotron radiation requires a description
of interactions~\cite{newjourn}.
But clearly the implications for
the cutoff energy for synchrotron radiation
should be kept in focus by future work on
(hopefully better developed) DSR phenomenology,
and it may provide other opportunities to distinguish between DSR scenarios
and Lorentz-symmetry-breaking scenarios: in fact,
a proper analysis of synchrotron radiation requires, besides suitable handling of dynamics/interactions,
also a description
of the laws of energy-momentum conservation~\cite{newjourn}.

\section{Some Other Results and Valuable Observations}
I have so far discussed only points that, according to my perspective, are the core
facts (and most ``virulently incorrect" myths) about DSR. While this is not intended as a well-rounded
review (and omissions, even of deserving works, will be inevitable),
especially for the benefit of DSR newcomers ({\it DSR aficionados} know
these and more) I think I should mention a few examples of other studies
that are representative of the range of scenarios that are under consideration
from a DSR perspective.

\subsection{A 2+1D DSR Theory?}
At least in some formulations of quantum gravity in 2+1 spacetime
dimensions a q-deformed deSitter symmetry Hopf algebra $SO(3,1)_q$
emerges~\cite{roche} for nonvanishing cosmological constant, and
the relation between the cosmological constant $\Lambda$ and
the $q$ deformation parameter takes the
form $\ln q \sim \sqrt{\Lambda} L_p$ for small $\Lambda$. It was observed in
Ref.~\cite{kodadsr} that
this relation $\ln q \sim \sqrt{\Lambda} L_p$
implies (in the sense of the "Hopf-algebra contractions" already considered
in Ref.~\cite{lukie91e92,firecontraction})
 that the $\Lambda \rightarrow 0$
limit is described by a $\kappa$-Poincar\'e Hopf algebra. However
the analysis only shows that the $\kappa$-Poincar\'e Hopf algebra
should have a role, without providing a fully physical picture.
But, since, as mentioned earlier in these notes, there
is some preliminary evidence that the
mathematics of the $\kappa$-Poincar\'e Hopf algebra might be
used to produce a DSR theory, the observation reported in
Ref.~\cite{kodadsr} generated some justifiable interest:
besides being an opportunity to perhaps find a genuine DSR toy theory,
this would also provide a striking picture for how the DSR framework
might emerge from a quantum-gravity theory.

It was then observed in Ref.~\cite{jurekkodadsr}
 that some aspects of the formulation of Matschull {\it et
al}~\cite{mats1,mats2,mats3} of classical gravity for point
particles in 2+1 dimensions are compatible with the DSR idea. The
key DSR-friendly ingredients are the presence of a maximum value
of mass and a description of energy-momentum space with ``deSitter
type" geometry (see later). However, several additional results must be
obtained in order to verify whether or not a DSR formulation is
possible. One key point is that Matschull {\it et al}
formulate~\cite{mats1,mats2} the theory by making explicit
reference to the frame of the center of mass of the multiparticle
system. It may therefore be illegitimate to assume that the
features that emerge from the analysis are observer independent.
Moreover, rather than a deformation of the
translation/rotation/boost classical symmetries of 2+1D space,
many aspects of the theory, because of an underlying conical
geometry, appear to be characterized by only two symmetries: a
rotation and a time translation. And this description in terms of
conical geometry is also closely related to the fact that
the ``observers at infinity" in the framework of Matschull {\it et al}
do not really decouple from the system under observation, and
therefore might not be good examples for testing the Relativity
Principle. Moreover, especially when considering particle
collisions, Refs.~\cite{mats1,mats2,mats3} appear to describe
frequently as total momentum of a multiparticle system simply the
sum of the individual momenta of the particles composing the
system, and, from a DSR perspective, such a linear-additivity law
would be incompatible~\cite{dsr1,dsr2} with a deformed dispersion relation.

I should also stress
that from a DSR perspective the 2+1 context might not provide the
correct intuition for the 3+1 context. A key difference
for what concerns the role of fundamental scales
originates from the fact that in 3+1
dimensions both the Planck length and the Planck energy are
related to the gravitational constant through the Planck constant
($L_p \equiv \sqrt{\hbar G/c^3}~,~~E_p \equiv \sqrt{\hbar c^5/
G}$), whereas in 2+1 dimensions the Planck energy is obtained only in
terms of the speed-of-light scale and
the gravitational constant: $E_p^{(2+1)} \equiv c^4/ G^{(2+1)}$.

So there are ``reasons for advancing cautiously", but
 still this should be considered
one of the most exciting developments for DSR research, and indeed
it continues to motivate several related studies (see, it e.g.,
Refs.~\cite{qg2DDSRfreidel,qg2DDSRoriti}).

\subsection{A Path for DSR in Loop Quantum Gravity?}
Another exciting possibility that has received some attention in the
DSR literature is the one of obtaining a DSR framework at some effective-theory
level of description of Loop Quantum Gravity.

An early suggestion of this possibility was formulated already in
Ref.~\cite{kodadsr}, taking as starting point the fact that the Loop Quantum
Gravity literature presents some support\cite{kodama,artem}
for the presence of a q-deformation of the deSitter symmetry
algebra when there is nonvanishing cosmological constant. As
discussed in Ref.~\cite{leekoda} (and preliminarily in
Ref.~\cite{kodadsr}), in the 3+1D context one expects a
renormalization of energy-momentum which is still not under
control, and for the analysis of Ref.~\cite{kodadsr} this
essentially translates into an inability to fully predict the relation
between the $q$-deformation parameter and the cosmological
constant, which in turn does not allow us to firmly establish
the potentialities of this line of analysis to produce a genuine
DSR framework.

While no fully robust derivation is available at present, the fact
that over these past few years other arguments and lines of analysis
have also suggested (see, {\it e.g.,},
Refs~\cite{lqgDSR1,lqgDSR2,carloDSR,leeDSRok}) that a DSR framework
might emerge at some effective-theory level of description of Loop
Quantum Gravity is certainly exciting.

\subsection{Maximum Momentum, Maximum Energy, Minimum Wavelength}
Going back to the more humble (but presently better controlled)
context of toy DSR test theories, it perhaps deserves stressing
(in spite of the limited scope and limited significance of such test theories)
that these test theories have been shown to accommodated rather
nicely some features that are appealing from the perspective
of a popular quantum-gravity intuition.
The toy test theory I discussed in the phenomenology section
would clearly predict a maximum Planckian ($1/\lambda$)
allowed value of spatial momentum.
And clearly a correspondingly structured test theory for frequencies/wavelengths
realizes a corresponding minimum-wavelength ($\lambda$) bound~\cite{dsr1,dsr2,dsrIJMPrev}.

Also much studied is a toy test theory that was first proposed
by Magueijo and Smolin~\cite{leedsr} in which
one arrives at a
maximum Planckian ($1/\lambda$)
allowed value of energy. This is a test theory with structure that is completely
analogous to the one of the DSR test theory I am here focusing on as illustrative
example, but based on the dispersion relation
\begin{equation}
\frac{E^2-\stackrel{\rightarrow}{p}^2}{(1-\lambda
E)^2}=\frac{m^2}{(1-\lambda m)^2} ~ \label{leejoao}
\end{equation}
rather than (\ref{dispKpoinnewAO}).

\subsection{Deformed Klein-Gordon/Dirac Equations}
Some progress has also been reached in formulating deformed
Klein-Gordon and Dirac equations in ways that would be consistent
with the structure of the toy DSR test theory I discussed
in the phenomenology section. This is done primarily
assuming that $\kappa$-Minkowski noncommutative spacetime
 provides an acceptable spacetime sector for our toy DSR
 test theory~\cite{dsrDIRAC},
 but interestingly it can also be done fully within ``energy-momentum space".
 In the case of the energy-momentum-space Dirac equation
 this latter possibility materializes~\cite{dsrDIRAC,dharamdsrdirac}
 in the following form
\begin{equation}\label{eqx}
  \left(\gamma^{\mu} {\cal D}_{\mu}(E,p,m;\lambda)-I\right)
  \psi(\overrightarrow{p})=0
\end{equation}
where
\begin{equation}\label{eqy}
 {\cal D}_{0} = \frac{e^{\lambda E}-\cosh\left(\lambda m \right)}
 {\sinh\left(\lambda m \right)} ~
\end{equation}
\begin{equation}\label{diki}
 {\cal D}_{j} = {p_j \over p} \frac{\left(2
e^{\lambda E}\left[\cosh\left(\lambda E \right)
 -\cosh\left(\lambda m \right)
\right]\right)^{\frac{1}{2}}}{\sinh\left(\lambda m \right)} ~
\end{equation}
 $I$ is the identity matrix,
and the $\gamma^{\mu}$ are the familiar ``$\gamma$ matrices".

\subsection{Looking beyond the ``Soccerball Problem"}
Within the specific setup of toy DSR test theories of the type I considered
in the phenomenology section, there is a natural issue to be considered,
which in the first days of development of such toy test
theories Kowalski-Glikman and I lightly referred to as ``the soccerball problem"
(a picturesque characterization still widely used, for example, at workshops where DSR researchers meet).
The point is that nonlinear deformations of the energy-momentum relation
are certainly phenomenologically acceptable for fundamental particles
(if the deformation scale is Planckian the effects are extremely small),
but clearly the same modifications of the energy-momentum relation
would be unacceptable for bodies with rest energy greater than the Planck
energy, such as the moon or a soccerball.

This issue remains somehow surrounded by a ``mystique" in some DSR circles,
but it is probably not very significant. Even within the confines
of that specific type of toy DSR test theories there probably is, as stressed already
in Refs.~\cite{dsr1,dsr2},  an easy
solution. In fact, another challenge for those test theories is the introduction
of a description of ``total momentum" for a body composed of many particles,
and the two difficulties might be linked:
it does not appear to be unlikely~\cite{dsr1,dsr2}
that the proper description of total momentum
might be such that the nonlinear properties attributed to individual particles are
automatically suppressed for a multiparticle body.

Another possibility is the one of
reformulating the test theory for frequencies/wavelengths rather
than for energy/momentum, in which case
the ``soccerball problem" essentially disappears~\cite{dsrIJMPrev}.

\subsection{Curvature in Energy-momentum Space}
While not adding to the scientific content of
toy DSR test theories of the type I considered
in the phenomenology section, it is conceptually intriguing and possibly valuable
to notice that the nonlinearities that these test theories implement can be viewed
as a manifestation of curvature in energy-momentum space. Indeed,
as primarily stressed by Kowalski-Glikman~\cite{lqgDSR2}
(applying in the DSR arena a line of
reasoning previously advocated, from a wider quantum-gravity perspective,
by Majid~\cite{majidbook,gacmaj}),
in such test theories the
energy-momentum variables can be viewed as the coordinates of a DeSitter-like
geometry.

Applying this viewpoint has proven valuable to quickly seeing
some properties of a given toy DSR test theory of the type I considered
in the phenomenology section. One must however keep in mind the physics
of the variables one is handling: it would be for example erroneous
 to assume that
diffeomorphism transformations of energy-momentum variables could be handled/viewed
in exactly the same way as diffeomorphism transformations of spacetime coordinates.

I should also stress that, while it might be valuable to view (when
possible) in terms of curvature in energy-momentum space a given
framework for which compatibility with the DSR principles has
already been independently established, the availability of a
natural-looking map from the energy-momentum variables to some
coordinates over a curved ({\it e.g.,} deSitter) geometry does not
suffice to guarantee the availability of a DSR formulation of the
relevant framework. In order to make this remark more concrete let
me propose a simple analogy. The propagation of light in a
water-pool is (to very good approximation) described by a dispersion
relation $E = \sqrt{c_{water}^2 p^2 + c_{water}^4 m^2}$ ($m=0$ for
photons), which of course allows a map from the energy-momentum
variables to some coordinates on a Minkowski geometry, but we know
that the scale $c_{water}$ is not observer independent, and the laws
of propagation of light in water do not admit a special-relativistic
formulation.

\subsection{A Rainbow Metric?}
In a sense similar to the usefulness of the observation concerning curvature
in energy-momentum space it may also be useful to look at toy DSR test
theories of the type I considered
in the phenomenology section as theories which, at least to some extent, are characterized
by an energy-dependent metric.
This was already the intuition behind some parts of my first DSR papers~\cite{dsr1,dsr2},
especially when I considered {\it gedanken}  procedures for the measurement
of distances in theories with observer-independent modifications of the dispersion relation
(because of the dependence on the energy of the probes found in the analysis of
measurements of the length of a physical/material object~\cite{dsr1,dsr2}).
More recently
Magueijo and Smolin~\cite{rainbowDSR} took these arguments to a higher level of abstraction,
and also introduced the well-chosen label ``rainbow metric".
The point being raised by Magueijo and Smolin could be very powerful,
since it would clearly be (if for no other reasons,
at least practically/computationally) advantageous to find a collection of
features of a DSR theory that could all be codified under the common unifying umbrella
of a given energy-dependent  metric.
However, in pursuing this objective it is of course necessary to proceed cautiously:
if, rather than rephrasing established DSR features of a given framework in terms
of a rainbow metric, one simply took a formal/technical approach,
introducing here and there (wherever feasible) an energy-dependent metric,
the end result might well not have DSR-compatible structure.
Of course, just like a postulate of ``curvature in energy-momentum space"
would not in itself suffice to qualify the corresponding framework as DSR-compatible,
the introduction of some energy-dependent metric does not automatically
lead to a DSR theory.
For example, the propagation of light in dispersive materials
often leads to a dispersion relation that could be formally
arranged in the form $p_\mu g(p_0;L_p)^{\mu \nu} p_\nu =0$, thereby
introducing an energy-dependent metric, but of course the presence
of a dispersion-inducing material actually selects a preferred frame, and
therefore is incompatible with any relativistic formulation.

Among the
examples of concepts that are discussed in the DSR literature and clearly do admit
rainbow-metric description let me also mention, in addition to
the aspects of distance/length measurement discussed in Ref.~\cite{dsr1,dsr2},
also the case of  modified
energy-momentum (dispersion) relations: indeed at least for massless particles
some of the modified energy-momentum relations that have most attracted attention
in the relevant literature, which can always
be cast in the form $f(p_\mu;L_p)=0$, could be rewritten (and be accordingly reinterpreted)
in the form $f(p_\mu;L_p)=p_\mu g(p_0;L_p)^{\mu \nu} p_\nu =0$.

A robust rainbow-metric description might be more challenging
(if at all available) in multiparticle contexts where
there is no obvious characteristic energy scale on which to link
the metric dependence. For formulas involving the energy-momentum
of several particles, with large hierarchies between the energies of different
particles in the system, it would seem that there might be no natural choice of
energy scale on which to anchor the rainbow metric.
An intriguing possibility in relation to these challenges
for the rainbow-metric description is perhaps found in the observations
I offered in Section~4: the concept of a energy-dependent metric is problematic
in a classical spacetime but may well be a valuable charterization
at the level of the quantum-spacetime picture.

\section{Some Recent Proposals}

\subsection{A Phase-Space-Algebra Approach}
Already in some of the first investigations of the DSR idea a few
authors (perhaps most notably Kowalski-Glikman (see, it e.g.,
Ref.~\cite{jurekDSRnew}) had explored the possibility of obtaining a
DSR-compatible framework by postulating a noncommutativity of
spacetime coordinates that would be merged within a single algebraic
structure with the symmetry-transformation generators, thereby
forming a 14-generator ``phase-space algebra". Recently the
development of a new approach essentially based on that perspective
was started in Refs.~\cite{gho1,gho2,gho3,gho4}, with the objective
of
 constructing point-particle
 models that possess a noncommutative (and non-canonical) simplectic
structure
and satisfy a modified dispersion relation, also hoping
that such a setup might facilitate the description of interactions
and the introduction of quantum properties for the point particles.
A valuable tool that has been developed~\cite{gho2} for this approach
is a map connecting the novel phase-space structures to ordinary
canonical phase space.
And among the features that already emerged from this
approach I should at least also mention an alternative derivation
of the Dirac equation (both in the case of the maximum-momentum toy DSR test theory
that I used as example in the phenomenology section, and for
the maximum-energy toy DSR test theory that was first considered by Magueijo and
Smolin), and a proposal of a Nambu-goto setup
which is expected~\cite{gho2} to be suitable for a DSR-compatible description of
point particles.

\subsection{Finsler Geometry}
>From various perspectives one can look at the structure
of the DSR principles as suggesting that spacetime should
be described in terms of some ``exotic" geometry,
and indeed, as mentioned, the majority of DSR studies assume
a noncommutative spacetime geometry.
There has been recently some exploration~\cite{dsrFinsl}
(also see Ref.~\cite{gho2})
from the DSR perspective of another candidate as exotic spacetime geometry:
Finsler geometry.

This idea is in its very early stage of exploration, and in particular
for the Finsler geometries considered in Refs.~\cite{dsrFinsl}
the presence of 10-generator Poincar\'e-like symmetries (a minimum prerequisite
to even contemplate a formalism as DSR candidate) has not yet
been established. It may be significant that
the Finsler line element is not invariant~\cite{mignemiFinsler}
under DSR-type transformations.

\subsection{A 5D Perspective}
Several DSR studies, while aiming for a proposal
for 4D physics, finds useful to adopt a formalism or a perspective which
is of 5D nature. For those exploring the possibility of obtaining a DSR-compatible
picture with theories formulated in $\kappa$-Minkowski noncommutative spacetime,
an invitation toward a 5D perspective comes from the fact that
for 4D $\kappa$-Minkowski spacetime it is not unnatural to consider a 5D differential
calculus~\cite{kowa5d,kappanoether5D}.
This mainly has its roots in the fact that
the $\kappa$-Poincar\'e Hopf algebra is most naturally introduced~\cite{lukie91e92}
as a contraction of the q-deSitter Hopf algebra (whose dual spacetime
picture is naturally described in terms of a 5D embedding environment).
And even some authors who are not advocating the noncommutative spacetime formulation
(and are probably unaware of the peculiarities of the 5D differential calculus)
have independently argued~\cite{dsr5Da,dsr5Db}
that it might be beneficial to set up the construction
of DSR theories from a sort of embedding-space 5D picture of spacetime variables and/or
energy-momentum variables.
It is probably fair to say that at present it is still unclear whether
the type of intuition generated by these 5D perspectives can be truly valuable,
but it is certainly striking that different areas of DSR research, working independently,
ended up advocating a 5D perspective.

\subsection{Everything Rainbow}
Taking off from the ``rainbow metric" perspective, which I briefly
discussed in the previous section, one can find arguments
to advocate also~\cite{hosse1} that other scales,
including the quantum-mechanics scale $\hbar$, should acquire an energy dependence.
This is too recent a proposal for me to comment robustly on it,
but I should stress that the challenges for interpretation of
some applications of the ``rainbow metric" that I mentioned in the previous
section of course also apply to this ``everything rainbow" approach,
and probably apply in more severe way.
 There are formulas in physics that refer simultaneously
 to the metric, to $\hbar$ and to a multitude of energy scales
 (formulas that apply to a large system, composed of many subsystems,
 for which one can meaningfully introduce a ``subsystem energy" concept),
 and in such cases it seems that one should find a perfect recipe
in order to constructively obtain an effective metric and
effective $\hbar$.

In partly related work~\cite{hosse2} it has been argued that rather
than concepts such as maximum space momentum, maximum energy and minimum wavelengths
(that I had so far mentioned in describing work on toy DSR test theories),
one could perhaps consider the development of a toy DSR theory in which
the new relativistic scale actually sets a maximum value for energy density.
This in an intriguing proposal, which might gain supporters if a
satisfactory relativistic formulation is found and explicitly articulated
in terms of operatively well-defined entities.
The construction of such a relativistic formulation may prove
however rather challenging, even more challenging than it has been
for the type of toy DSR test theories I considered in the phenomenology section.
In fact, the laws of transformation of energy density should require
from the onset a description of symmetry transformations that acts
simultaneously on energy-momentum variables (energy) and spacetime variables (the volume
where the relevant energy is ``contained" which is of course needed for the energy-density
considerations).

\section{Closing Remarks}
Doubly-Special Relativity is maturing quickly, as a result of the
interest it is attracting from various research groups, each
bringing its relevant expertise to the programme.
A key factor for the future success of this large multi-perspective effort
is the adoption of a common language, used to give and gain full
access to the different bits of
progress achieved toward the development of the DSR idea.
I have here attempted to contribute to this goal, advocating
the usefulness of characterizing results from a physics-content
perspective (rather than technical/mathematical),
consistently with the spirit of my first DSR papers~\cite{dsr1,dsr2}.

Clearly the key objective at this point should be the one of
constructing a full theory (with spacetime observables, energy-momentum observables,
frequencies, wavelengths, cross sections...) compatible with the DSR requirements.
I have mentioned here some partial results which might suggest that this
ambitious objective is now not far. In the meantime the use of ``toy DSR test theories"
(in the sense I have here clarified) can be valuable, both to develop some intuition
for what type of effects could be predicted within a DSR framework, and to
anchor on some formulas (of however limited scope) the debate on conceptual
aspects of DSR research.

\section*{Acknowledgments}
My work was supported in part by grant RFP2-08-02 from The Foundational
Questions Institute (fqxi.org).
The preparation of this manuscript was diluted over a long time, and
conversations with several colleagues were valuable for my work,
in one way or another. I am particularly grateful to
F. Fiore, S. Ghosh,
G. Gubitosi,
S. Hossenfelder, U. Jacob, J. Kowalski-Glikman,
A. Marcian\'o, M. Matassa, F. Mercati, T. Piran, G. Rosati,
and L. Smolin.

\newpage
%==========================================================
% Back Matter (References and Notes)
%----------------------------------------------------------
% Style and layout of the references
\bibliographystyle{mdpi}
\makeatletter
\renewcommand\@biblabel[1]{#1. }
\makeatother
%----------------------------------------------------------
% Use the following option to include external BibTeX files:
%\bibliography{template}

\begin{thebibliography}{000}

%\begin{thebibliography}{000}

\bibitem{dsr1} Amelino-Camelia, G.
Relativity in space-times with short-distance
structure governed by an observer-independent (Planckian)
length scale.
gr-qc/0012051,
%Int.~J.~Mod.~Phys.~{\bf D11}, 35 (2002) 35--60;
{\em Int.~J.~Mod.~Phys.}~{\bf 2002}, {\em D11},  35--60.

\bibitem{dsr2} Amelino-Camelia, G.
Testable scenario for Relativity
with minimum length.
hep-th/0012238,
 {\em Phys.~Lett.}~{\bf 2001} {\em B510}
255--263.
%{\bibit Phys.~Lett.}~{\bibbf B510}, 255 (2001).

\bibitem{jurekdsr} Kowalski-Glikman, J.
Observer independent quantum of mass.
%hep-th/0102098, Phys.~Lett.~A286 (2001) 391--394.
hep-th/0102098,
{\em Phys.~Lett.}~ {\bf 2001}, {\em A286}, 391--394.

\bibitem{gacRoxJurek} Bruno, N.R.; Amelino-Camelia, G.; Kowalski-Glikman, J.
Deformed boost transformations that saturate at the Planck scale.
hep-th/0107039,
{\em Phys.~Lett.}~{\bf 2001}, {\em B522}, 133--138.
%DEFORMED BOOST TRANSFORMATIONS THAT SATURATE AT THE PLANCK SCALE.

\bibitem{cosmodsr} Alexander, S.; Brandenberger, R.; Magueijo, J.
Noncommutative inflation.
hep-th/0108190,
{\em Phys.~Rev.}~{\bf 2003}, {\em D67}, 081301.
%NONCOMMUTATIVE INFLATION.

\bibitem{leedsr}  Magueijo, J.; Smolin, L.
Lorentz invariance with an invariant energy scale.
hep-th/0112090,
{\em  Phys.~Rev.~Lett.}~{\bf 2002}, {\em 88}, 190403.

\bibitem{frandar} Amelino-Camelia, G.;
Benedetti, D.; D'Andrea, F.; Procaccini, A. Comparison of relativity
theories with observer independent scales of both velocity and
length/mass.
hep-th/0201245,
{\em Class.~Quant.~Grav.}~{\bf 2003}, {\em 20}, 5353--5370.

\bibitem{lukiedsr}
Lukierski J.; A.~Nowicki, A. Doubly special relativity versus kappa
deformation of relativistic kinematics.
hep-th/0203065,
{\em Int.~J.~Mod.~Phys.} {\bf 2003}, {\em A18}, 7--18.

\bibitem{jurekDSRnew} Kowalski-Glikman, J.; Nowak, S.
Noncommutative space-time of doubly special relativity theories.
%ONCOMMUTATIVE SPACE-TIME OF DOUBLY SPECIAL RELATIVITY THEORIES.
hep-th/0204245,
{\em Int.~J.~Mod.~Phys.}~{\bf 2003}, {\em D12}, 299--316.

\bibitem{judesvisser} Judes, S.; Visser, M.
Conservation laws in Doubly special relativity.
gr-qc/0205067, {\em Phys.Rev.} {\bf 2003}, {\em D68}, 045001.

\bibitem{dsrDIRAC} Agostini, A.; Amelino-Camelia, G.; Arzano, M.
%DIRAC SPINORS    FOR DOUBLY SPECIAL RELATIVITY.
Dirac spinors for doubly special relativity and kappa Minkowski
noncummutative space-time. gr-qc/0207003,
{\em Class.~Quant.~Grav.}~{\bf 2004},
{\em 21}, 2179--2202.

\bibitem{dharamdsrdirac} Ahluwalia, D.V.
Fermions, bosons, and locality in special relativity with two
invariant scales.
gr-qc/0207004v4.

\bibitem{dsrnature} Amelino-Camelia, G.
%DOUBLY SPECIAL RELATIVITY.
%gr-qc/0207049, Nature 418 (2002) 34--35.
Relativity: Special treatment.
gr-qc/0207049,
{\em Nature} {\bf 2002}, {\em 418}, 34--35.

\bibitem{dsrlodz} Rembielinski, J.; Smolinski,  K.A.
Unphysical Predictions of Some Doubly Special Relativity Theories.
hep-th/0207031,
{\em Bull.~Soc.~Sci.~Lett.~Lodz} {\bf 2003}, {\em 53}, 57--63.

\bibitem{leeDSRprd} Magueijo, J.; Smolin, L.
Generalized Lorentz invariance with an invariant energy scale.
%GENERALIZED LORENTZ INVARIANCE WITH AN INVARIANT ENERGY SCALE.
gr-qc/0207085,
{\em Phys.Rev.} {\bf 2003}, {\em D67}, 044017.

\bibitem{rossano} Bruno, N.R.
Group of boost and rotation transformations with two observer independent scales.
%GROUP OF BOOST AND ROTATION TRANSFORMATIONS WITH TWO OBSERVER INDEPENDENT SCALES
gr-qc/0207076,
{\em Phys.~Lett.} {\bf 2002}, {\em B547}, 109--115.

\bibitem{dsrgzk} Amelino-Camelia, G.
Kinematical solution of the UHE-cosmic-ray puzzle without a preferred class of inertial observers.
%Published in Int.J.Mod.Phys.D12:1211-1226,2003.
astro-ph/0209232,
{\em Int.~J.~Mod.~Phys.}~{\bf 2003}, {\em D12} 1211--1226.

\bibitem{feoli} Feoli, A.
Maximal acceleration or maximal accelerations?
gr-qc/0210038,
{\em Int.~J.~Mod.~Phys.}~{\bf 2003}, {\em D12}, 271--280.

\bibitem{dsrIJMPrev} Amelino-Camelia, G.
 Doubly special relativity: First results and key open problems.
%Published in Int.J.Mod.Phys.D11:1643,2002.
gr-qc/0210063,
 %DOUBLY SPECIAL RELATIVITY: FIRST RESULTS AND KEY OPEN PROBLEMS.
 {\em Int.~J.~Mod.~Phys.} {\bf 2002}, {\em D11}, 1643--1669.

\bibitem{dsrchak} Chakrabarti, A.
Nonlinear transforms of momenta and Planck scale limit.
hep-th/0211214,
%Published in J.Math.Phys.44:3800-3808,2003.
{\em J.~Math.~Phys.}~{\bf 2003}, {\em 44} 3800.

\bibitem{dsrBlautKowa} Blaut, A.; Daszkiewicz, M.; Kowalski-Glikman, J.
Doubly Special Relativity with Light-Cone Deformation.
hep-th/0302157,
{\em Mod.~Phys.~Lett.}~{\bf 2003}, {\em A18}, 1711--1719.

\bibitem{rainbowDSR} Magueijo, J.; Smolin, L.
Gravity's rainbow.
%Published in Class.Quant.Grav.21:1725-1736,2004.
gr-qc/0305055,
{\em Class.~Quant.~Grav.}~{\bf 2004}, {\em 21}, 1725--1736.

\bibitem{maguespace} Kimberly, D.; Magueijo, J.; Medeiros, J.
Nonlinear relativity in position space.
gr-qc/0303067,
{\em Phys.~Rev.}~{\bf 2004}, {\em D70}, 084007.

\bibitem{balROXher} Ballestero, A.; Bruno, N.R.; Herranz, F.J.
A New doubly special relativity theory from a quantum conformal algebra.
hep-th/0305033,
{\em J.~Phys.}~{\bf 2003}, {\em A36}, 10493--10503.

\bibitem{kodadsr} Amelino-Camelia, G.; Smolin, L.; Starodubtsev, A.
Quantum symmetry, the cosmological constant and Planck scale phenomenology.
hep-th/0306134,
%Class.Quant.Grav.21:3095-3110,2004
{\em Class.~Quant.~Grav.}~{\bf 2004}, {\em 21}, 3095--3110.

\bibitem{jurekkodadsr} Freidel, L.; Kowalski-Glikman, J.; Smolin, L.
2+1 gravity and doubly special relativity.
%2+1 GRAVITY AND DOUBLY SPECIAL RELATIVITY.
hep-th/0307085,
{\em Phys.Rev.} {\bf 2004}, {\em  D69}, 044001.

\bibitem{sethdsr} Heyman, D.; Hinteleitner, F.; Major, S.
On reaction thresholds in DSR theories.
gr-qc/0312089,
 {\em Phys.~Rev.}~{\bf 2004}, {\em D69}, 105016.

\bibitem{dsrphen} Amelino-Camelia, G.; Kowalski-Glikman, J.;
Mandanici, G.; Procaccini, A.
Phenomenology of doubly special relativity.
%PHENOMENOLOGY OF DOUBLY SPECIAL RELATIVITY.
gr-qc/0312124,
{\em Int.~J.~Mod.~Phys.}~{\bf 2005},
{\em A20}, 6007.

\bibitem{stringydsr} Magueijo, J.; Smolin, L.
String theories with deformed energy momentum relations, and a possible nontachyonic bosonic string.
hep-th/0401087,
{\em Phys.~Rev.~} {\bf 2005}, {\em D71}, 026010.

\bibitem{mg10qg5} Amelino-Camelia, G.
Some encouraging and some cautionary remarks on doubly special
relativity in quantum gravity.
gr-qc/0402092.

\bibitem{mignemidsr} Mignemi, S.
Hamiltonian formalism and space-time symmetries in generic DSR models.
gr-qc/0403038,
{\em Int. J. Mod. Phys.} {\bf 2006}, {\em D15}, 925--936.

\bibitem{dsrlngsold} Aloisio, R.; Carmona, J.M.; Cortes, J.L.; Galante, A.;
Grillo, A.F.; Mendez, F. Particle and antiparticle sectors in DSR1
and kappa-Minkowski space-time.
hep-th/0404111,
 {\em JHEP} {\bf 2004}, {\em 0405}, 028.

\bibitem{dsroriti} Livine, E.R.; Oriti, D.
About Lorentz invariance in a discrete quantum setting.
gr-qc/0405085,
{\em JHEP} {\bf 2004}, {\em 0406}, 050.

\bibitem{corgamboa} Cortes, J.L.; Gamboa, J.
 Quantum uncertainty in doubly special relativity.
hep-th/0405285,
{\em Phys.~Rev.}~{\bf 2005}, {\em D71}, 065015.

\bibitem{tsr} Kowalski-Glikman, J.; Smolin, L.
Triply special relativity.
hep-th/0406276,
{\em Phys.~Rev.}~{\bf 2004}, {\em D70}, 065020.

\bibitem{qg2DDSRfreidel} Freidel, L.; Livine, E.R.
Ponzano-Regge model revisited III: Feynman diagrams and effective field theory.
hep-th/0502106,
{\em Class.~Quant.~Grav.}~{\bf 2006}, {\em 23}, 2021--2062.

\bibitem{qg2DDSRoriti} Livine, E.R.; Oriti, D.
Coherent states for 3-D deformed special relativity: Semi-classical points in a quantum flat spacetime.
hep-th/0509192,
{\em JHEP} {\bf 2006}, {\em 0511}, 050.

\bibitem{lqgDSR1} Smolin, L.
Falsifiable predictions from semiclassical quantum gravity.
hep-th/0501091,
 {\em Nucl.~Phys.}~{\bf 2006}, {\em B742}, 142--157.

\bibitem{lqgDSR2} Imilkowska, K.; Kowalski-Glikman, J.
Doubly special relativity as a limit of gravity.
gr-qc/0506084,
{\em Lect.~Notes Phys.}~{\bf 2006}, {\em 702}, 279--298.

\bibitem{gho1} Ghosh, S.; Pal, P.
 hep-th/0502192,
Kappa-Minkowski spacetime through exotic 'oscillator'. {\em
Phys.~Lett.}~{\bf 2005}, {\em B618}, 243--251.

\bibitem{gho2} Ghosh, S.; Pal, P.
 Deformed Special Relativity and Deformed Symmetries in a Canonical Framework.
hep-th/0702159,
{\em Phys.~Rev.}~{\bf 2007}, {\em D75}, 105021.

\bibitem{gho3} Ghosh, S.
DSR relativistic particle in a Lagrangian formulation and
non-commutative spacetime: A gauge independent analysis.
 {\em Phys.~Lett.}~{\bf 2007}, {\em B648}, 262--265.

\bibitem{gho4} Gosselin, P.; Berard, A.; Mohrbach, H.; Ghosh, S.
Berry phase effects in the dynamics of Dirac electrons in doubly special relativity framework.
arXiv:0709.0579,
{\em Phys.~Lett.}~{\bf 2008}, {\em B660}, 267--274.

\bibitem{leivadsr} Calisto, H.; Leiva, C.
Generalized commutation relations and DSR theories, a close relationship.
hep-th/0509227,
{\em Int.~J.~Mod.~Phys.}~{\bf 2007}, {\em D16},  927--940.

\bibitem{lngsdsr} Aloisio, R.; Galante, A.; Grillo, A.;  Liberati, S.;
Luzio, E.; Mendez, F. Deformed special relativity as an effective
theory of measurements on quantum gravitational backgrounds.
gr-qc/0511031,
{\em Phys.~Rev.}~{\bf 2006}, {\em D73}, 045020.

\bibitem{freepartyDSR} Girelli, F.; Konopka, T.; Kowalski-Glikman, J.;
Livine, E.R. The Free particle in deformed special relativity.
hep-th/0512107,
 {\em Phys.~Rev.}~{\bf 2006}, {\em D73}, 045009.

\bibitem{ling} Ling, Y.; Li, X.;
Zhang, H.-b. Thermodynamics of modified black holes from gravity's
rainbow.
gr-qc/0512084,
{\em Mod.~Phys.~Lett.}~{\bf 2007}, {\em A22}, 2749--2756.

\bibitem{konopdsr} Konopka, T.
A Field theory model with a new Lorentz-invariant energy scale.
hep-th/0601030,
{\em Mod.~Phys.~Lett.}~{\bf 2008}, {\em A23}, 319--335.

\bibitem{irandsr} Jafari, N.; Shariati, A.
Doubly special relativity: A New relativity or not?
gr-qc/0602075,
{\em AIP Conf.~Proc.}~{\bf 2006}, {\em 841}, 462--465.

\bibitem{dsrDas} Das, A.; Kong, O.C.W.
Physics of quantum relativity through a linear realization.
gr-qc/0603114,
{\em Phys.~Rev.}~{\bf 2006}, {\em D73}, 124029.

\bibitem{dsrSaoPaulo} Aldrovandi, R.; Beltran Almeida, J.P.;
Pereira, J.G.
de Sitter special relativity.
gr-qc/0606122,
{\em Class. Quant. Grav.} {\bf 2007}, {\em 24}, 1385-1404.

\bibitem{dsrAurelio0607} Aloisio, R.; Galante, A.; Grillo, A.F.; Liberati, S.;
Luzio, E.; Mendez, F. Modified special relativity on a fluctuating
spacetime.
gr-qc/0607024,
{\em Phys.~Rev.}~{\bf 2006}, {\em D74}, 085017.

\bibitem{dsrFinsl} Girelli, F.; Liberati, S.; Sindoni, L.
Planck-scale modified dispersion relations and Finsler geometry.
gr-qc/0611024,
{\em Phys.~Rev.}~{\bf 2007}, {\em D75}, 064015.

\bibitem{hosse1} Hossenfelder, S.
Deformed special relativity in position space.
gr-qc/0612167,
{\em Phys.~Lett.}~{\bf 2007}, {\em B649}, 310--316.

\bibitem{hosse2} Hossenfelder, S.
Multi-Particle states in deformed special relativity.
hep-th/0702016,
{\em Phys.~Rev.}~{\bf 2007}, {\em D75}, 105005.

\bibitem{dsr5Da} Freidel, L.; Girelli, F.; Livine, E.R.
The relativistic particle: Dirac observables and feynman propagator.
hep-th/0701113,
{\em Phys.~Rev.}~{\bf 2007}, {\em D75}, 105016.

\bibitem{dsr5Db} Girelli, F.; Livine, E.R.
Non-Commutativity of effective space-time coordinates and the
minimal length.
arXiv:0708.3813.

\bibitem{dsrMadrid} Galan, P.; Mena Marugan, G.A.
Canonical realizations of doubly special relativity.
gr-qc/0702027,
{\em Int. J. Mod. Phys.} {\bf 2007}, {\em D16}, 1133--1147.

\bibitem{dsrZagreb} Kresic-Juric, S.; Meljanac, S.; Stojic, M.
Covariant realizations of kappa-deformed space.
hep-th/0702215,
{\em Eur. Phys. J.} {\bf 2007}, {\em C51}, 229--240.

\bibitem{dsrBeijing} Liu, C.-Z.; Zhu, J.-Y.
Asymptotic quasinormal modes of scalar field in a gravity's rainbow.
gr-qc/0703058.
{\em Chin.~Phys.} {\bf 2009}, {\em B18}, 4161--4168.

\bibitem{mignemiFinsler} Mignemi, S.
 Doubly special relativity and Finsler geometry.
arXiv:0704.1728,
{\em Phys.~Rev.}~{\bf 2007}, {\em D76}, 047702.

\bibitem{dsrchaiho} Kim, H.-C.; Rim, C.; Yee, J.H.
Blackbody radiation in kappa-Minkowski spacetime.
arXiv:0705.4628,
{\em Phys.~Rev.}~{\bf 2007}, {\em D76}, 105012.

\bibitem{dsrHinter} Hinterleitner, F.
Remarks on DSR and gravity.
arXiv:0706.0471,
{\em Class.~Quant.~Grav.} {\bf 2008}, {\em 25}, 075018

\bibitem{dsrGianluca} Mandanici, G.
Undeformed (additive) energy conservation law in Doubly Special
Relativity. arXiv:0707.3700, {\em Mod. Phys. Lett.} {\bf 2009} {\em A24}, 739--745.

\bibitem{grbgac} Amelino-Camelia, G.; Ellis, J.; Mavromatos, N.E.; Nanopoulos, D.V.;
Sarkar, S. Tests of quantum gravity from observations of
$\gamma$-ray bursts.
astro-ph/9712103,
%Nature {393} (1998) 763--765.
{\em Nature} ~{\bf 1998}, {\em  393}, 763--765.
%{\bibit Nature} {\bibbf 393}, 763 (1998).

\bibitem{gampul} Gambini, R.; Pullin, J.
% NONSTANDARD OPTICS FROM QUANTUM SPACE-TIME.
Nonstandard optics from quantum space-time.
 gr-qc/9809038,
{\em Phys.~Rev.}~{\bf 1999}, {\em D59}, 124021.

\bibitem{kifune} Kifune, T.
Invariance violation extends the cosmic ray horizon?
astro-ph/9904164,
%{\bibit Astrophys.~J.~Lett.}~{\bibbf 518}, L21 (1999).
{\em Astrophys.~J.~Lett.} ~{\bf 1999}, {\em 518}, L21--L24.

\bibitem{mexweave} Alfaro, J.; Morales-Tecotl, H.A.; Urrutia, L.F.
%QUANTUM GRAVITY CORRECTIONS TO NEUTRINO PROPAGATION.
%Phys.Rev.Lett.84:2318-2321,2000
Quantum gravity corrections to neutrino propagation.
%{\bibit Phys.~Rev.~Lett.}~{\bibbf 84}, 2318 (2000).
gr-qc/9909079,
{\em Phys.~Rev.~Lett.} ~{\bf 2000}, {\em 84}, 2318--2321.

\bibitem{biller} Schaefer B.E.;
Severe Limits on Variations of the Speed of Light with Frequency
%Phys.~Rev~Lett.~82 (1999) 4964-4966.
{\em Phys.~Rev.~Lett.} {\bf 1999}, {\em 82}, 4964--4966;
Biller, S.D. et al.
Limits to Quantum Gravity Effects on Energy Dependence of the Speed
of Light from Observations of TeV Flares in Active Galaxies.
%%{\bibit Phys.~Rev.~Lett.}~{\bibbf 83}, 2108 (1999).
{\em Phys.~Rev.~Lett.} ~{\bf 1999}, {\em 83}, 2108--2011.

\bibitem{ita} Aloisio, R.; Blasi, P.; Ghia, P.L.; Grillo, A.F.
Probing The Structure of Space-Time with Cosmic Rays.
astro-ph/0001258,
{\em Phys.~Rev.}~{\bf 2000}, {\em D62}, 053010.

\bibitem{aus} Protheroe, R.J.; Meyer, H.
An infrared background TeV gamma ray crisis?
astro-ph/0005349,
{\em Phys.~Lett.}~{\bf 2000},\linebreak\ {\em B493}, 1--6.
%Phys.~Lett.~B493 (2000) 1-6.

\bibitem{gactp}  Amelino-Camelia, G.; Piran, T.
Planck scale deformation of Lorentz symmetry as a solution to the
UHECR and the TeV gamma paradoxes.
astro-ph/0008107,
{\em Phys.~Rev.} ~{\bf 2001}, {\em D64}, 036005.

\bibitem{tedOLDgood} Jacobson, T.; Liberati, S.; Mattingly, D.
%TEV ASTROPHYSICS CONSTRAINTS ON PLANCK SCALE LORENTZ VIOLATION.
TeV astrophysics constraints on Planck scale Lorentz violation.
hep-ph/0112207,
{\em Phys.~Rev.}~{\bf 2002}, {\em D66},  081302.

\bibitem{gacpion} Amelino-Camelia, G.
Space-time quantum solves three experimental paradoxes.
gr-qc/0107086,
 {\em Phys.~Lett.}~{\bf 2002}, {\em B528}, 181--187.
%{\it Phys.~Lett.}~B {\bf 528}, 181--187 (2002).

\bibitem{seth} Konopka, T.J.; Major, S.A.
hep-ph/0201184,
Observational limits on quantum geometry effects. {\em New
J.~Phys.}\linebreak\ {\bf 2002}, {\em 4}, 57.

\bibitem{kappanoether} Agostini, A.; Amelino-Camelia, G.;
Arzano, M.; Marcian\`o, A.; Tacchi, R.A. Generalizing the Noether
theorem for Hopf-algebra spacetime symmetries.
hep-th/0607221,
{\em Mod.~Phys.~Lett.}~{\bf 2007}, {\em A22}, 1779--1786.

\bibitem{thetanoether} Amelino-Camelia, G.; Briscese, F.; Gubitosi, G.; Marcian\`o, A.;
Martinetti, P.; Mercati, F. Noether analysis of the twisted Hopf
symmetries of canonical noncommutative spacetimes.
arXiv:0709.4600,
{\em Phys.~Rev.} {\bf 2008}, {\em D78} 025005.

\bibitem{lukie91e92} Lukierski, J.; Ruegg, H.;  Nowicki, A.; Tolstoi, V.N.
Q deformation of Poincare algebra. {\em Phys. Lett.} {\bf 1991},
{\em B264}, 331--338.

\bibitem{majrue} Majid, S.; Ruegg, H.
Bicrossproduct structure of kappa {P}oincar{\'{e}} group
and noncommutative geometry.
%{\bibit Phys.~Lett.}~{\bibbf B334}, 348 (1994).
{\em Phys.~Lett.}~{\bf 1994}, {\em B334}, 348--354.

\bibitem{kpoinap} Lukierski, J.; Ruegg, H.; Zakrzewski, W.J.
 Classical and quantum-mechanics
of free $\kappa$-relativistic systems.
hep-th/9312153,
{\em Ann. Phys.}~{\bf 1995},
{\em 243}, 90--116.
%Ann. Phys. {\bf 243} (1995) 90--116.
%{\bibit Ann.~Phys.}~{\bibbf 243}, 90 (1995).

\bibitem{kpoinnogroup} Lukierski, J.; Ruegg, H.; Ruhl, W.
From kappa Poincare algebra to kappa Lorentz quasigroup: A Deformation of relativistic symmetry.
%FROM KAPPA POINCARE ALGEBRA TO KAPPA LORENTZ QUASIGROUP
{\em Phys.~Lett.} ~{\bf 1993}, {\em B313}, 357-366.
%Phys.~Lett.~B313 (1993) 357--366.
%\bibitem{kpoinnogroup} J.~Lukierski, H.~Ruegg and W.~Ruhl,
%%FROM KAPPA POINCARE ALGEBRA TO KAPPA LORENTZ QUASIGROUP
%Phys.~Lett.~B313 (1993) 357.
%%Phys.~Lett.~B313 (1993) 357--366.

\bibitem{batalin} Batalin, I.A.
Quasigroup construction and first class constraints.
%J.~Math.~Phys.~22 (1981) 1837-1850.
{\em J.~Math.~Phys.}~{\bf 1981}, {\em 22}, 1837--1850.

\bibitem{fockbook} Fock V. {\em The theory of space-time and gravitation};
Pergamon Press: Oxford, UK, 1964.
[the relevant analysis is in the appendix]

\bibitem{cardone} Cardone, F.;  Marrani, A.; Mignani, R.
Boosts in an Arbitrary Direction and Maximal Causal Velocities in a
Deformed Minkowski Space.
 {\em Found.~Phys.~Lett.}~{\bf 2003}, {\em 16}, 163--181.

\bibitem{areaNEWpap} Amelino-Camelia, G.
On the fate of Lorentz symmetry in loop quantum gravity and
noncommutative space-times.
arXiv:gr-qc/0205125.

\bibitem{snyder} Snyder, H.S.;
Quantized Space-Time. {\em Phys.~Rev.}~{\bf 1947}, {\em 71}, 38--41.

\bibitem{kempmang} Kempf, A.; Mangano, G.; Mann, R.B.
%HILBERT SPACE REPRESENTATION OF THE MINIMAL LENGTH UNCERTAINTY RELATION.
%Published in Phys.Rev.D52:1108--1118,1995
%e-Print Archive: hep-th/9412167
Hilbert space representation of the minimal length uncertainty
relation.
{\em Phys.~Rev.}~{\bf 1995}, {\em D52} 1108--1118.

\bibitem{dadebro} Ahluwalia, D.V.
Wave particle duality at the Planck scale: Freezing of neutrino
oscillations.
gr-qc/0002005, {\em Phys. Lett.} {\bf 2000}, {\em A275}, 31--35.
%WAVE-PARTICLE DUALITY AT THE PLANCK SCALE: FREEZING OF NEUTRINO
%OSCILLATIONS.
%Published in Phys.Lett.A275:31-35,2000
%QUANTUM VIOLATION OF THE EQUIVALENCE PRINCIPLE AND GRAVITATIONALLY
%MODIFIED DE BROGLIE RELATION.
%By D.V. Ahluwalia (Zacatecas U.). ISGBG-02, Feb 2000. 9pp.
%Published in Phys.Lett.A275:31-35,2000
%e-Print Archive: gr-qc/0002005

\bibitem{bignapapPRD} Amelino-Camelia, G.
Gravity-wave interferometers as probes of a
low-energy effective quantum gravity.
gr-qc/9903080,
{\em Phys.~Rev.}~{\bf 2000}, {\em D62}, 024015.

\bibitem{aadluna} Agostini, A.; Amelino-Camelia, G.; D'Andrea, F.
Hopf algebra description of noncommutative-spacetime symmetries.
%Int. J. Mod. Phys. A19 5187-5219 (2004).
hep-th/0306013,
{\em Int.~J.~Mod.~Phys.}~{\bf 2004}, {\em A19}, 5187--5219.
%%CITATION = HEP-TH 0306013;%%

\bibitem{nopureboost}
Amelino-Camelia, G.; Gubitosi,  G.; Marcian\`o, A.; Martinetti, P.;
Mercati, F. A no-pure-boost uncertainty principle from spacetime
noncommutativity.
 arXiv:0707.1863,
%Phys.Lett.B671:298-302,2009.
{\em Phys.~Lett.}~{\bf 2000}, {\em B671}, 298--302.

\bibitem{kowa5d}
Freidel, L.; Kowalski-Glikman, J.; Nowak, S. Field theory on
$\kappa$--Minkowski space revisited: Noether charges and breaking of
Lorentz symmetry.
hep-th/0706.3658,
%Int.J.Mod.Phys.A23:2687-2718,2008.
{\em Int.~J.~Mod.~Phys.}~{\bf 2008}, {\em A23},  2687--2718.

\bibitem{kappanoether5D} Amelino-Camelia, G.; Marcian\`o, A.; Pranzetti, D.
On the 5D differential calculus and translation transformations
in 4D kappa-Minkowski noncommutative spacetime.
%Int.J.Mod.Phys.A24:5445--5463,2009
arXiv:0709.2063,
{\em Int.~J.~Mod.~Phys.}~{\bf 2009}, {\em A24},  5445--5463.

\bibitem{doplich1}
Doplicher S.; Fredenhagen, K.;  Roberts, J.E.; The Quantum structure
of space-time at the Planck scale and quantum fields.
hep-th/0303037,
{\em Commun.~Math.~Phys.}~{\bf 1995}, {\em 172}, 187--220.

\bibitem{doplich2} Bahns D.; Doplicher, S.; Fredenhagen, K.;  Piacitelli, G.; Field
theory on noncommutative spacetimes: Quasiplanar Wick products.
hep-th/0408204, {\em Phys.~Rev.}~{\bf 2005}, {\em D71}, 025022.

\bibitem{jaconature} Jacobson, T.; Liberati, S.; Mattingly, D.
A strong astrophysical constraint on the violation of
special relativity by quantum gravity.
%A STRONG ASTROPHYSICAL CONSTRAINT ON THE VIOLATION OF SPECIAL RELATIVITY
%BY QUANTUM GRAVITY.
%astro-ph/0212190, Nature 424 (2003) 1019--1021
astro-ph/0212190,
%astro-ph/0212190v2,
{\em Nature}~{\bf 2003}, {\em 424}, 1019--1021.
%Nature {424}, 1019 (2003).

\bibitem{jackson} Jackson, J.D. {\em Classical Electrodynamics},
3rd ed.; J.~Wiley \& Sons, New York, NY, USA, 1999.

\bibitem{newjourn} Amelino-Camelia, G.
%IMPROVED LIMIT ON QUANTUM SPACE-TIME MODIFICATIONS OF LORENTZ SYMMETRY FROM
%OBSERVATIONS OF GAMMA-RAY BLAZARS.
Phenomenology of Planck-scale Lorentz-symmetry test theories.
gr-qc/0212002,
{\em New J.~Phys.} {\bf 2004}, {\em 6}, 188.

\bibitem{roche} Buffenoir, E.; Noui, K.; Roche, P.
Hamiltonian quantization of Chern-Simons theory with SL(2,C) group.
hep-th/0202121,
{\em Class.~Quant.~Grav.} {\bf 2002}, {\em 19}, 4953--5015.

\bibitem{firecontraction} Celeghini, E.; Giachetti, R.; Sorace, E.; Tarlini, M.
Three-dimensional quantum groups from contractions of SU(2)q.
{\em J. Math. Phys.} {\bf 1990}, {\em 31}, 2548-2551.
%E. Celeghini, R. Giacchetti, E. Sorace and M. Tarlini, J. Math. Phys. 31 (1990) 2548

\bibitem{mats1} Matschull, H.-J.; Welling, M.
%QUANTUM MECHANICS OF A POINT PARTICLE IN (2+1)-DIMENSIONAL GRAVITY.
Quantum mechanics of a point particle in (2+1)-dimensional gravity.
gr-qc/9708054,
{\em Class.~Quant.~Grav.} {\bf 1998}, {\em 15}, 2981--3030.

\bibitem{mats2} Louko, J.; Matschull, H.-J.
%(2+1)-DIMENSIONAL EINSTEIN-KEPLER PROBLEM IN THE CENTER-OF-MASS FRAME.
(2+1)-dimensional Einstein-Kepler problem in the center-of-mass
frame.
gr-qc/9908025,
 {\em Class.~Quant.~Grav.} {\bf 2000}, {\em 17}, 1847--1873.

\bibitem{mats3} Matschull, H.-J.
%THE PHASE SPACE STRUCTURE OF MULTI PARTICLE MODELS IN 2+1 GRAVITY.
The Phase space structure of multi
particle models in 2+1 gravity.
gr-qc/0103084,
{\em Class.~Quant.~Grav.} {\bf 2001}, {\em 18}, 3497--3560.

\bibitem{kodama} Kodama, H.
Holomorphic wave function of the Universe.
%{\it Holomorphic Wave Function Of The Universe},
{\em Phys.~Rev.} {\bf 1990}, {\em D42}, 2548-2565.

\bibitem{artem} Starodubtsev, A.
Topological excitations around the vacuum of quantum gravity I: The
symmetries of the vacuum.
hep-th/0306135.

\bibitem{leekoda} Smolin, L. Quantum gravity with a positive cosmological constant.
hep-th/0209079.

\bibitem{carloDSR} Rovelli, C.
A Note on DSR.
arXiv:0808.3505.

\bibitem{leeDSRok} Smolin, L.
Could deformed special relativity naturally arise from the
semiclassical limit of quantum gravity?
arXiv:0808.3765.

\bibitem{majidbook} Majid, S.
{\em Foundations of Quantum Group Theory}; Cambridge University
Press: Cambridge, UK, 1995.

\bibitem{gacmaj} Amelino-Camelia, G.; Majid, S.
Waves on noncommutative spacetime and gamma ray bursts.
hep-th/9907110,
%Int.~J. Mod.~Phys.~A15 (2000) 4301--4324.
{\em Int.~J. Mod.~Phys.} {\bf 2000}, {\em A15}, 4301--4324.


\end{thebibliography}
%----------------------------------------------------------

\end{document}